\documentclass[aps,prb,reprint,amsmath,amssymb]{revtex4-2}
\usepackage{inputenc}
\usepackage[english]{babel}
\addto\captionsenglish{}
\usepackage{graphicx,color}
\usepackage{wrapfig}
\usepackage{microtype}
\usepackage{enumitem}
\usepackage{dsfont}
\usepackage{siunitx}
\usepackage[caption=false]{subfig}
\usepackage{textcomp}
\usepackage{stmaryrd}
\usepackage[dvipsnames]{xcolor}
\usepackage[
  colorlinks=true,
  citecolor=TealBlue,
  linkcolor=MidnightBlue,
  urlcolor=RubineRed
]{hyperref}
\usepackage{titlesec}
\usepackage{gensymb}
\usepackage{braket}
\usepackage{mathtools} 
\usepackage{placeins}

\usepackage{tikz}
\usetikzlibrary{shapes}
\usepackage{xcolor}

\usepackage{cancel}

\definecolor{darkgray}{HTML}{595959}

\newcommand{\bhexagon}{\mathord{\raisebox{0.6pt}{\tikz{\node[draw,scale=.8,regular polygon, regular polygon sides=6,rotate=90,fill=darkgray,darkgray](){};}}}}

\def\vk{{\boldsymbol{k}}}

\def\vr{{\boldsymbol{r}}}
\def\vq{{\boldsymbol{q}}}

\def\vb{{\boldsymbol{b}}}

\def\vd{{\boldsymbol{d}}}
\def\vl{{\boldsymbol{l}}}

\def\ve{{\boldsymbol{e}}}

\def\vr{{\boldsymbol{r}}}

\def\vK{{\boldsymbol{K}}}
\def\vG{{\boldsymbol{G}}}

\newcommand\varpm{\mathbin{\vcenter{\hbox{\oalign{\hfil$\scriptstyle+$\hfil\cr\noalign{\kern-.3ex}$\scriptscriptstyle({-})$\cr}}}}}

\hypersetup{pdfencoding=auto,hidelinks}

\begin{document}

\title{Quasi-flat and exact flat bands in patterned two-dimensional electron gases and honeycomb wire networks}

\author{Chunxiao Liu \textsuperscript{1,2}}
\email{chunxiao.liu@universite-paris-saclay.fr}

\author{Igor Pishniak\textsuperscript{3}}

\author{Benoît Douçot \textsuperscript{4}}

\author{Jérôme Cayssol \textsuperscript{3}}
\email{jerome.cayssol@u-bordeaux.fr}

\affiliation{\textsuperscript{1}Department of Physics, University of California, Berkeley, California 94720, USA}

\affiliation{\textsuperscript{2}Université Paris-Saclay, CNRS, Laboratoire de Physique des Solides, 91405 Orsay, France}

\affiliation{\textsuperscript{3}Univ. Bordeaux, CNRS, LOMA, UMR 5798, F-33400, Talence, France}

\affiliation{\textsuperscript{4}Sorbonne Université, CNRS, Laboratoire de Physique Théorique des Hautes Énergies, LPTHE, F-75005 Paris, France}

\date{\today}

\begin{abstract}

We investigate the formation of flat bands in a two-dimensional electron gas patterned by a periodic triangular array of repulsive antidots. In the high-barrier, narrow-trench regime, the spectrum develops {\it kagome-like} band triplets, each comprising a nearly flat band and two dispersive bands, with wave functions concentrated along the potential-free trenches. The two lowest energy bands have a different character with {\it graphene-like} dispersion and wave functions localized near the junction centers. The appearance of the trench-concentrated nearly flat bands further motivates a honeycomb quantum graph model as an effective description of the narrow-trench limit. In the single-channel graph, destructive interference produces exact flat bands with compact localized eigenstates protected by the $D_3$ symmetry. Their energies obey a quantization condition determined by propagation along the links and scattering at the junctions. We then extend the analysis to a multichannel quantum graph and derive the most general junction scattering matrix compatible with $D_3$ symmetry. Even- and odd-parity transverse channels contain equivalent symmetry sectors, and $D_3$ symmetry therefore allows them to mix. This mode conversion can disrupt the destructive interference responsible for flat bands and generically gives those bands a finite width. Exact flatness is nevertheless retained when only one transverse-parity sector is active, or when an additional symmetry forbids mixing between the two sectors. These results connect the continuum antidot model to its quantum-graph limit and identify transverse-mode mixing as a mechanism that limits flatness in multichannel networks.

\end{abstract}

\maketitle

\tableofcontents

\section{Introduction} 

In periodic media, the behavior of excitations is governed by the underlying band structure of the energy spectrum \cite{Girvin_Yang_2019}. While materials generically exhibit dispersive bands separated by gaps, the possibility to engineer flat bands (FBs) allows to quench the kinetic energy of electrons and therefore to enhance the effects of Coulomb interaction \cite{Derzhko2015,Checkelsky2024} and disorder \cite{Bilitewski2018}. Although the associated ordinary velocity cancels, FBs lead to rich phenomena due to the quantum geometry of Bloch and Wannier states \cite{Torma2023}. Besides the Landau levels of the quantum Hall effect \cite{Klitzing1980}, FBs and their associated compact localized states (CLS) can arise from destructive interference of hopping processes in frustrated lattices like the kagome, Lieb or dice lattices \cite{Sutherland1986,Mielke1991,Mielke1992,Tasaki1992,Vidal1998}, without requiring any magnetic field. Recently certain real materials, layered or even 3D, were shown to host nearly FBs  \cite{Lin2018,Kang_NatCom2020,Kang2020,Huang2022,Hu2022,Lou2022}, thereby motivating systematic search for FBs in tight-binding models \cite{Hase2018,Calugaru2022,Regnault2022,Sabri2023,Neves2024,Li2025}. 

Recent advances in nanofabrication have enabled the realization of highly tunable solid-state artificial crystals through engineered superlattice potentials \cite{Du2018,Wang2018,Wang2023,Wang2026}. In moiré systems \cite{bistritzer2011moire,Balents2020}, the superlattice electrostatic potential arises from the relative twist between atomically thin layers and can give rise to nearly flat bands at selected magic angles. The symmetry of the external potential is set by the moiré pattern and its strength is fixed by the interlayer coupling. An alternative approach is to engineer the periodic potential by patterning a metallic top gate in the vicinity of a high-mobility two-dimensional electron gas (2DEG) \cite{Gibertini2009,Wang2023,Wang2026} or by etching directly the host semiconductor \cite{Du2018,Wang2018}. These strategies offer considerable flexibility in designing the superlattice geometry and enables independent control of the potential landscape and carrier density through electrostatic gating. Typical superlattice periods are on the order of 100 nm, corresponding to bandwidths of approximately 1 meV. The residual disorder in the 2DEG can be drastically reduced using high-mobility GaAs/AlGaAs heterostructures or encapsulated graphene. Therefore these platforms offer a promising route toward the realization of FBs and the exploration of interaction-driven correlated phases \cite{Wang2023,Wang2026}. 

In contrast to the exact flat bands of the kagome tight-binding model \cite{Bergman2008}, only nearly flat bands emerges in 2D semiconducting heterostructures and in moiré systems at magic angles \cite{Balents2020}. Nevertheless, in some limits, such patterned 2DEGs may resemble a 1D network of well defined trenches, which can be described by a 1D quantum graph model. It turns out that the purely 1D (single channel) honeycomb quantum graph (HQG) model has exact flat bands \cite{Kuchment2007,Deoliveira2022} which are reminiscent of the kagome tight-binding model \cite{Bergman2008}. Therefore in the crossover from the 2D antidot triangular lattice towards the 1D HQG, the quasi-flat bands become exact ones. Understanding the mechanism of this phenomenon could help optimizing designs of patterned 2DEG to achieve flatter bands. The 1D HQG limit is also expected to be realized in some (non-magical-angle) moir\'e materials \cite{Efimkin2018,Pal2019,Debeule2023,Zandvliet2025} where chiral edge states can realize such an HQG. Historically, continuous differential operators acting along one-dimensional QGs \cite{berkolaiko2013,Akkermans2000} have been used to model aromatic molecules \cite{Pauling1936,Griffith1952}, arrays of 1D metallic wires \cite{Coulson1954,Kazymyrenko2005}, percolation in various contexts \cite{deGennes1981,Fink1982,Alexander1983,Rammal1983,Chalker1988}, quantum chaos \cite{Kottos1997,Kottos1999,Kottos2000,Anantharaman2021}, and even finite-order approximations to fractals \cite{Kempkes2019,Bercioux2019}.

In this work, we investigate the emergence of nearly flat bands (FBs) in a high-mobility two-dimensional electron gas (2DEG) patterned by a triangular array of hexagonal antidots, and explore the crossover from the two-dimensional system to the one-dimensional quantum graph limit. In our two-dimensional model, electrons are excluded from hexagonal obstacles and confined to the intervening trenches, which form a honeycomb wire network. For sufficiently strong confinement, the spectrum comprises two {\it graphene-like} bands at lowest energies and higher {\it kagome-like} band triplets containing nearly flat and dispersive bands. For narrow trenches, a single-channel honeycomb quantum graph provides an effective description of the {\it kagome-like} bands. In this idealized model, the flat bands are exact, i.e. their bandwidth is rigorously zero. The resulting spectrum consists of repeated groups of two dispersive bands plus one singular flat band \cite{Rhim2019}, reminiscent of the band structure of the kagome lattice \cite{Bergman2008}.
To characterize this limiting regime, we introduce a single-channel quantum graph model. We then extend this description to a multichannel quantum wire network with additional propagating transverse modes. Symmetry-allowed scattering between modes of opposite transverse parity generically broadens the graph's flat bands. This identifies even--odd transverse-mode conversion as a mechanism that can limit flatness in finite-width networks such as those formed by the antidot potential.

The paper is organized as follows. In Sec.~\ref{sec:2DEGantidots}, we investigate the spectral properties of a two-dimensional electron gas (2DEG) patterned by a periodic triangular array of hexagonal antidots with an emphasis on the nearly flat bands emerging in a wide range of parameters when the superlattice potential is strong enough. In Section~\ref{sec:singlechannel}, it is shown that the single-channel honeycomb quantum graph, which a limit of the 2D antidot model, generically harbors exact flat bands protected by the $D_3$ symmetry. In Section~\ref{sec:multichannel}, we study the multichannel HQG as an intermediate model between the patterned 2DEG and the single-channel graph. This allows to show that the destruction of the graph exact FBs originates from the elastic coupling between transverse channels with different parities. Section~\ref{sec:conclusion} summarizes our results and discusses possible experimental implementations.

\section{Patterned 2DEG with an antidot potential
\label{sec:2DEGantidots}}
Artificial electronic structures can be engineered by patterning a high-mobility two-dimensional electron gas (2DEG) using electrostatic gates \cite{Wang2026} or etched obstacles \cite{Du2018,Wang2018}. The repulsive antidot potential depletes the electron density in selected regions while preserving conduction in the surrounding two-dimensional electron gas. Recent experiments have demonstrated the potential of this approach by employing triangular arrays of holes or pillars to engineer an artificial band structure in a GaAs/AlGaAs heterostructure \cite{Wang2018,Wang2026}. Remarkably, the same device was shown to exhibit {\it graphene-like} dispersion at low energies and {\it kagome-like} dispersion at higher energies \cite{Wang2026}. Analogous structures can be realized on metallic surfaces by positioning repulsive molecules with a scanning tunneling microscope \cite{Slot2017,Kempkes2019,Bercioux2019}. All these available modern platforms motivate the continuous 2D antidot model considered in this section, consisting of a triangular lattice of repulsive hexagonal antidots separated by potential-free trenches. We investigate its low-energy spectrum as a function of the antidot barrier height $V_0$ and trench width $w$ and the favorable conditions for observing the narrower {\it kagome-like} nearly flat bands. We first solve a 2D antidot model using plane wave expansion (Secs.~\ref{Sec:II.A} and \ref{Sec:II.B}). We then describe the {\it graphene-like} bands using a two-orbital model (Sec.~\ref{sec:2DEGantidotsBoundStates}) and examine the width of the lowest nearly flat {\it kagome-like} band as the potential profile varies (Sec.~\ref{sec:2DEGantidotsTrenchStates}).

\subsection{Model and numerical formulation}
\label{Sec:II.A}

\begin{figure}[t]
    \centering
    \includegraphics[width=\columnwidth]{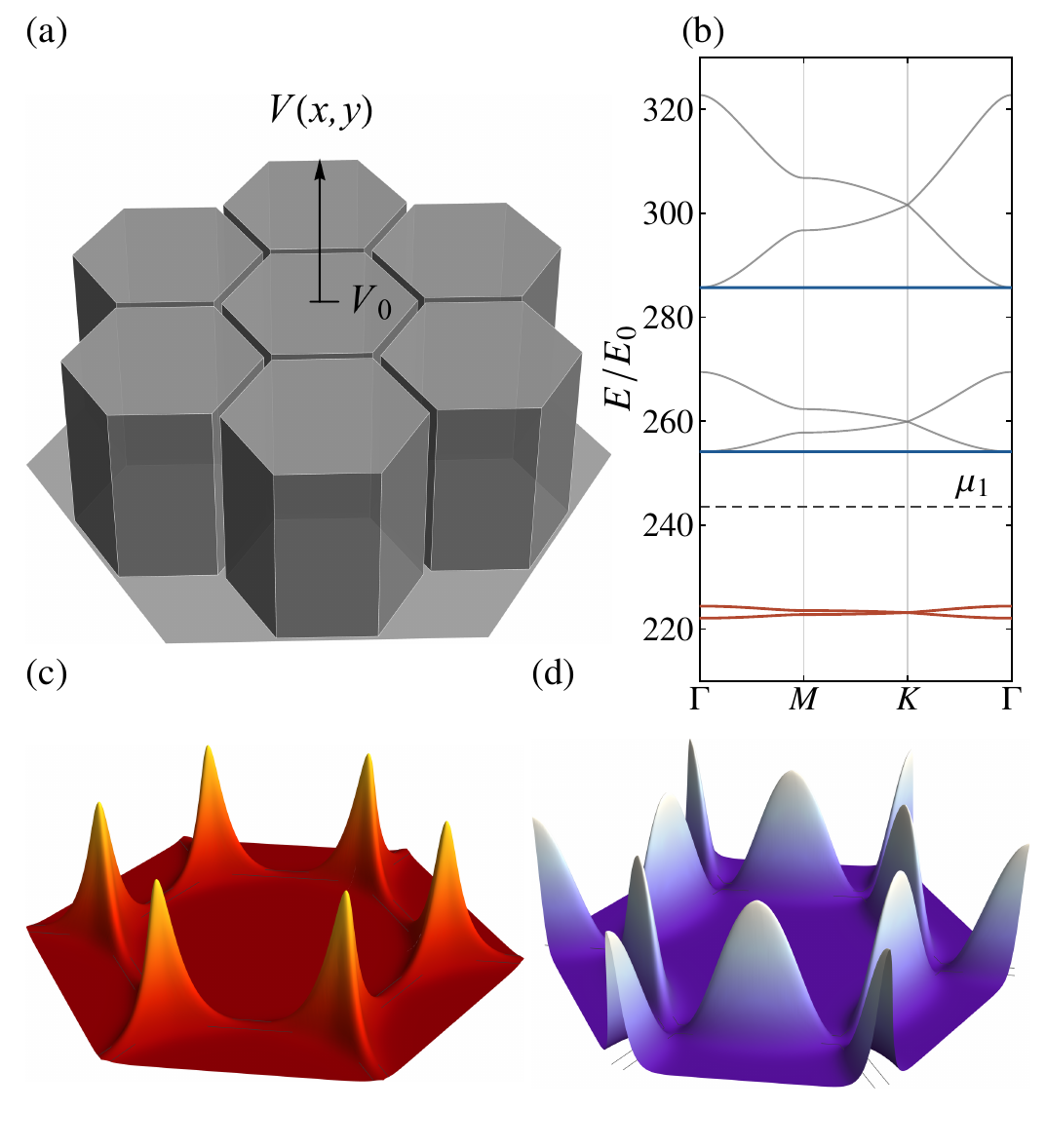}
    \caption{Superlattice electrostatic potential $V(x,y)$ and probability density $|\Psi|^2$ for the low-energy states.
    (a) Hexagonal antidot regions with potential $V_0>0$, separated by potential-free
    trenches of width $w$.
    (b) The lowest eight bands along the high-symmetry path
    $\Gamma$--$M$--$K$--$\Gamma$. The two lowest bands ($n=1,2$, red) are {\it graphene-like}, with wave functions concentrated near the junction centers. The higher triplets ($n=3,4,5$ and $n=6,7,8$) are {\it kagome-like}, with weight concentrated along the trenches. The bands $n=3$ and $n=6$ (blue) are nearly flat; the residual width of band $n=3$ is shown in Fig.~\ref{fig:antidot_bandwidth_map}. The dashed horizontal line marks the lowest straight-trench threshold $\mu_1(V_0,w)$, used as the energy origin in Fig.~\ref{fig:antidot_graph_crossover}.
    (c) Normalized probability density of
    the {\it graphene-like} bands ($n=1,2$) at $\Gamma$, obtained by summing over the densities of each band.
    (d) Normalized, equal-weight incoherent average of the lowest nearly flat {\it kagome-like} band ($n=3$)
    densities at the three symmetry-related $M$ points. In each of panels (c) and (d), surface height and color both represent the probability density, rescaled to unit maximum separately in each panel. 
    All panels use $V_0/E_0=400$ and $w/(\sqrt{3}a_0)=0.05$, which corresponds to a strong potential regime with well defined trenches.}
    \label{fig:antidot_model_states}
\end{figure}

We consider a free electron described by the single-particle Schrödinger Hamiltonian 
\begin{equation}\label{eq:antidot_hamiltonian}
 H=-\frac{\hbar^2}{2m}\nabla^2+V(\vr),
\end{equation}
where $m$ is the effective mass of electrons in the host 2DEG and $\vr=(x,y)$ the cartesian coordinates within the plane. The potential $V(\vr)$ is created by antidots whose centers form a triangular
Bravais lattice generated by the vectors :
\begin{equation}\label{defe1e2}
 \ve_1=\left(\sqrt3 ,3\right) \frac{a_0}{2}
 \hspace{3mm} {\rm and}  \hspace{3mm} 
 \ve_2=\left(-\sqrt3,3\right) \frac{a_0}{2} \, \, .
\end{equation}
Our model for the electrostatic superlattice potential reads
\begin{equation}\label{eq:Kronig_Penney_V}
V(\vr)=\left\{
\begin{array}{ll}
V_0,&\displaystyle
\vr\in\bigcup_{m,n\in\mathbb Z}\bhexagon_{(m,n),h},\\[2mm]
0,&\text{otherwise,}
\end{array}
\right.
\end{equation}
where $V_0 >0$ (repulsive antidots) and the symbol $\bhexagon_{(m,n),h}$ denotes the regular hexagon of side length $h$ centered
at $m\ve_1+n\ve_2$ (Fig.~\ref{fig:antidot_model_states}(a)). The distance $a_0$ between neighboring threefold Y junction centers is on the order of $\sim 10^2$ nm. The potential-free separation between adjacent antidots is $w$. The side length of the hexagons is therefore $h=a_0-w/\sqrt{3}$. A natural energy scale is provided by $E_0=\hbar^2/(2ma_0^2)$, which is $5.7 \times 10^{-2}$ meV using the effective mass of the electron in GaAs and $a_0=100$ nm.

For any Bloch state $\psi_\vk(\vr)=e^{i\vk\cdot\vr}u_\vk(\vr)$, the cell-periodic part $u_\vk(\vr)$ obeys the Schr\"odinger equation
\begin{equation}\label{KPHp}
 \left[\frac{\hbar^2}{2m}(-i\nabla+\vk)^2+V(\vr)\right]u_\vk(\vr)
 =E(\vk)u_\vk(\vr) \,\, .
\end{equation}
The $\vk$-dependent Bloch Hamiltonian reads
\begin{equation}\label{h_mat}
\begin{aligned}
H_{m,n;m',n'}(\vk)={}&
\frac{\hbar^2}{2m}|m\vb_1+n\vb_2+\vk|^2\delta_{m,m'}\delta_{n,n'}\\
&+V_{m,n;m',n'} \, 
\end{aligned}
\end{equation}
in the plane-wave basis $\ket{m,n}=A^{-1/2}e^{i(m\vb_1+n\vb_2)\cdot\vr}$, where $\vb_1$ and $\vb_2$ are the reciprocal primitive vectors, and $A$ the unit-cell area. The potential matrix is obtained from the Fourier transform of one hexagonal antidot potential. We then calculate the low-energy Bloch spectrum. The calculation details are given in Appendix~\ref{app:antidot_numerics}.

At each momentum, we order the eigenvalues
$E_1(\vk)\leq E_2(\vk)\leq\cdots$. We display the spectra along paths connecting the high-symmetry points $\Gamma$, $M$, and $K$, as shown in Fig.~\ref{fig:antidot_model_states}(b).
To quantify the dispersion of these bands and its dependence on $w$ and $V_0$, we define the bandwidth of band $n$ over the Brillouin zone as
\begin{equation}\label{eq:Wn_fullBZ}
 W_n=\max_{\vk\in\mathrm{BZ}}E_n(\vk)
     -\min_{\vk\in\mathrm{BZ}}E_n(\vk) \, \, .
\end{equation}

\subsection{\textit{Graphene-like} and \textit{kagome-like} dispersions}\label{Sec:II.B}

Fig.~\ref{fig:antidot_model_states}(b) shows a representative low-energy
spectrum of the patterned 2DEG described by Eq.~(\ref{eq:antidot_hamiltonian}). Since the two lowest bands ($n=1,2$) have Dirac crossings at $K$ and $K'$ and a van Hove singularity at $M$, we refer to this pair as the {\it graphene-like} bands. The next three bands ($n=3,4,5$) form a {\it kagome-like} triplet: one nearly flat band touches one of its two dispersive partners at $\Gamma$. A similar coexistence of {\it graphene-like} and {\it kagome-like} minibands was observed in Ref.~\cite{Wang2026}, using a potential profile different from the idealized one in Eq.~\eqref{eq:Kronig_Penney_V}. At higher energies, one observes the repetition (up to a deformation due to the quadratic dispersion of the host 2D) of the {\it kagome-like} trio ($n=6$--$8$,...).  

The two band families also have distinct spatial profiles. The {\it graphene-like} wave functions are concentrated near the Y junction centers [Fig.~\ref{fig:antidot_model_states}(c)], while the {\it kagome-like} wave functions have weight along the potential-free trenches surrounding the antidots, as illustrated for band $n=3$ in Fig.~\ref{fig:antidot_model_states}(d). For this illustration, we use its densities at the three symmetry-related $M$ points, where the band is separated from its neighboring bands. We average the densities with equal weight to restore threefold symmetry.
The dispersive bands $n=4,5$ also have Dirac crossings at $K$.

\begin{figure}[t]
    \centering
    \includegraphics[width=\columnwidth]{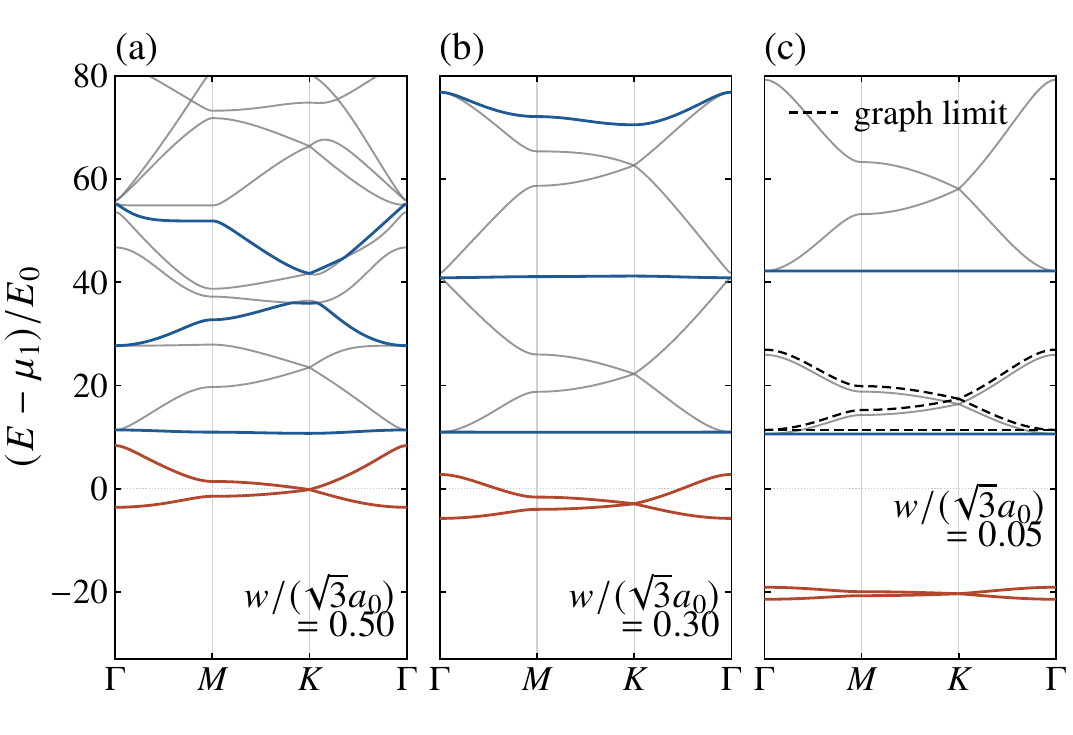}
    \caption{Evolution of the continuum antidot spectrum with trench width at
    fixed potential height $V_0/E_0=400$.  Panels (a)--(c) show the bands along
    $\Gamma$--$M$--$K$--$\Gamma$ for
    $w/(\sqrt{3}a_0)=0.50$, $0.30$, and $0.05$, respectively.  The energy is
    measured relative to $\mu_1$, the lowest transverse eigenenergy of an
    isolated straight trench. The {\it graphene-like} bands ($n=1,2$) are red. Bands $n=3,6,9$, the nearly flat members of the {\it kagome-like} triplets in the narrow-trench regime of panel (c), are blue wherever they enter the displayed energy window; all other bands are gray. Every band
    intersecting the displayed energy window is included.  The dashed curves
    in panel (c) are the single-channel quantum-graph result derived in
    Sec.~\ref{sec:singlechannel}; they are calculated independently of the
    periodic continuum bands. The threshold $\mu_1(V_0,w)$ is calculated
    for each width with the same barrier height, its calculation being described
    in Appendix~\ref{app:antidot_numerics}. Its value for panel (c) is marked
    in Fig.~\ref{fig:antidot_model_states}(b). The displayed range $-33\leq (E-\mu_1)/E_0\leq80$ spans approximately $6.4$ meV using the effective mass of the electron in GaAs and $a_0=100$ nm. The potentiel height $V_0=400 E_0 \simeq 22.8$ meV corresponds to the strong potential regime of experiments like \cite{Wang2026}.}
    \label{fig:antidot_graph_crossover}
\end{figure}

We now follow the evolution of these {\it graphene-like} and {\it kagome-like} bands as the trench width increases. The nearly flat members of the {\it kagome-like} triplets acquire dispersion (Fig.~\ref{fig:antidot_graph_crossover}). At $w/(\sqrt{3}a_0)=0.30$, band $n=9$ starts to disperse, while the {\it graphene-like} bands move closer to band $n=3$. For even shallower trenches with $w/(\sqrt{3}a_0)=0.50$, both the formerly quasi-flat bands $n=6$ and the $n=9$ band become substantially dispersive. Intuitively, one can already anticipate that in the limit of narrow trenches and high antidot barriers the 2D model maps to a 1D honeycomb quantum graph (HQG) limit (Fig.~\ref{fig:conventions}). In Sec.~\ref{sec:singlechannel}, we will show that this HQG always gives rise to flat bands embedded within those trios of {\it kagome-like} bands, for any node scattering, as long as the $D_3$ vertex symmetry is fulfilled.

To understand the contrasting bandwidths of the {\it graphene-like} bands ($n=1,2$) and the lowest nearly flat {\it kagome-like} band ($n=3$), we next compare their dependence on $V_0$ and $w$. Figs.~\ref{fig:antidot_junction_splitting} and \ref{fig:antidot_bandwidth_map} show, respectively, the splitting between the {\it graphene-like} bands at $\Gamma$ and the width of band $n=3$ over the $(w,V_0)$ parameter plane. We first discuss the two lowest bands in Sec.~\ref{sec:2DEGantidotsBoundStates} and then the third band in Sec.~\ref{sec:2DEGantidotsTrenchStates}. 

\subsection{\textit{Graphene-like} lowest bands $E_{1,2}(\vk)$ \label{sec:2DEGantidotsBoundStates}}
\begin{figure}[t]
    \centering
    \includegraphics[width=\columnwidth]{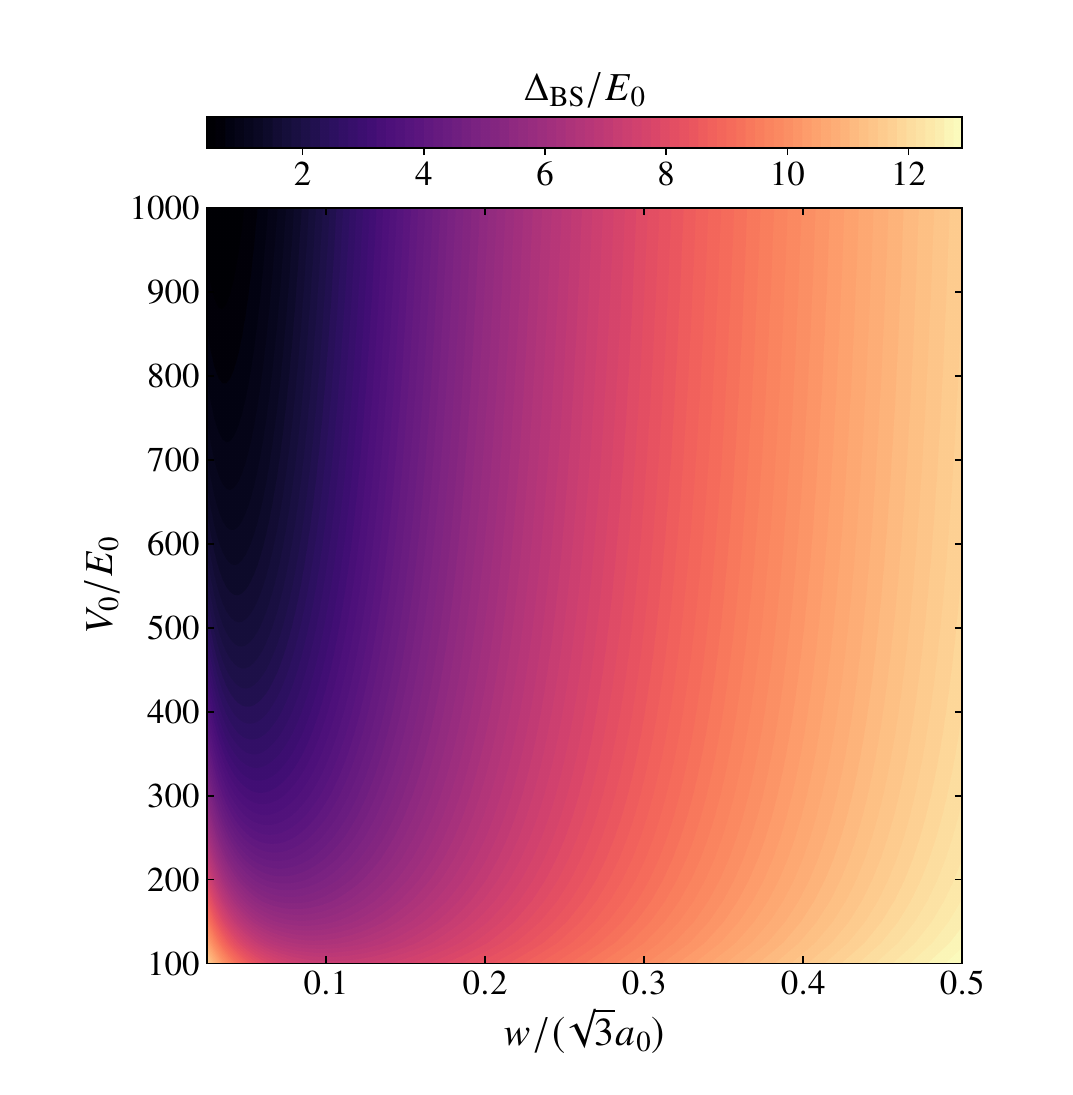}
    \caption{Splitting $\Delta_{\mathrm{BS}}=E_2(\Gamma)-E_1(\Gamma)$
    between the two {\it graphene-like} bands as a function of
    antidot strength $V_0$ and trench width $w$.
    In the nearest-neighbor two-orbital description,
    $\Delta_{\mathrm{BS}}=6 t_J$ and $W_1=W_2=3 t_J$.}
    \label{fig:antidot_junction_splitting}
\end{figure}
\begin{figure}[t]
    \centering
    \includegraphics[width=\columnwidth]{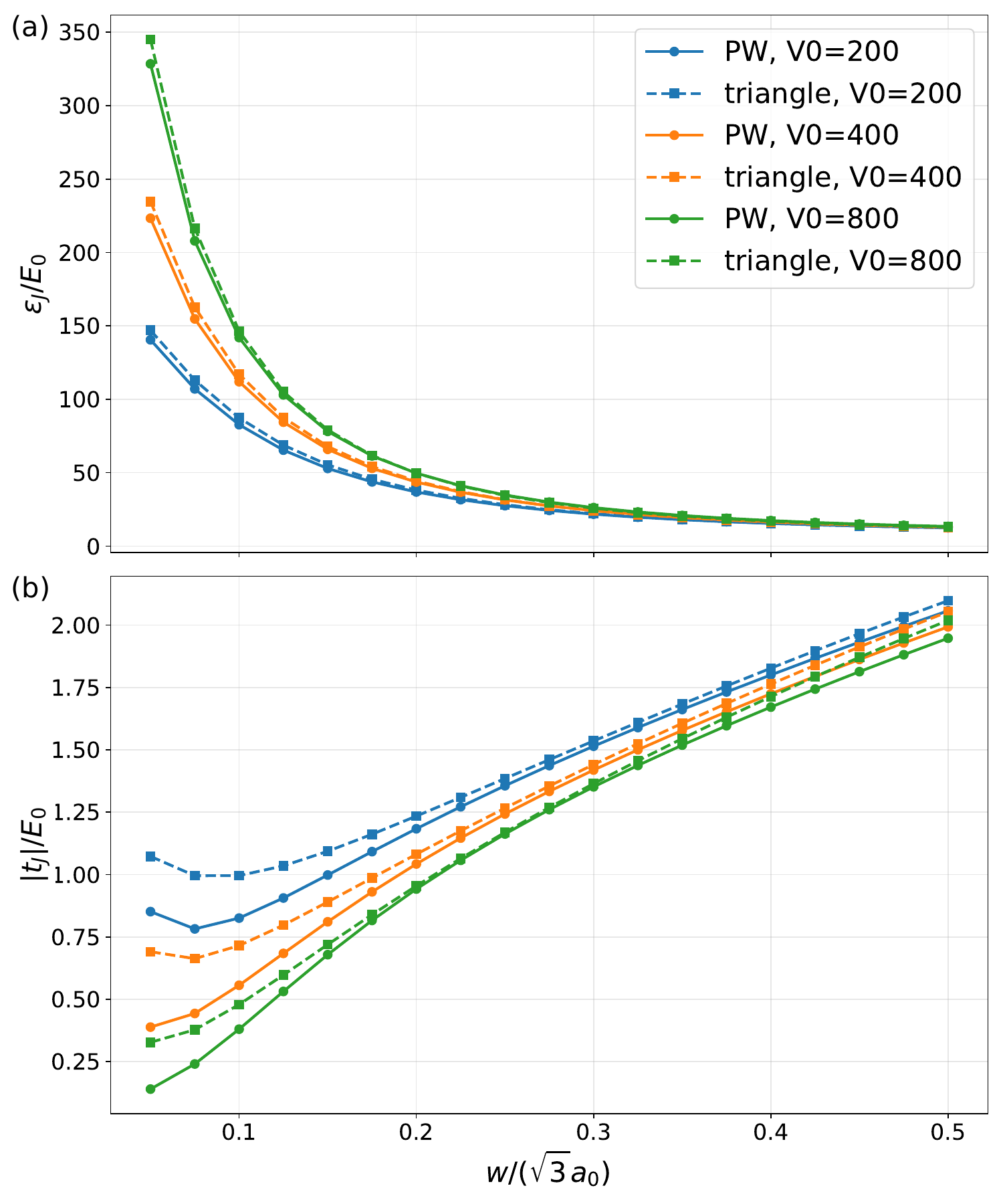}
    \caption{Comparison of the triangular-junction model with the full periodic two-dimensional calculation. (a) Junction energy $\epsilon_J$ and (b) nearest-neighbor hopping $|t_J|$ as functions of the normalized trench width $w/(\sqrt{3}a_0)$, for $V_0/E_0=200$, $400$, and $800$. The periodic values are extracted from the plane-wave band structure, the analytical values follow from the triangular-junction model.}
    \label{fig:triangle-junction}
\end{figure}

In order to model the {\it graphene-like} bands, one junction-centered orbital is placed on each site of the honeycomb lattice and only nearest-neighbor hopping is retained. These orbitals play the role of the carbon $p_z$ orbitals in the graphene tight-binding model. The dispersion reads
\begin{equation}\label{eq:junction_two_orbital}
 E_{n=1,2}(\vk)=\epsilon_J
 \mp t_J\left|1+e^{i\vk\cdot\ve_1}
                       +e^{i\vk\cdot\ve_2}\right| \,\, ,
\end{equation}
where $t_J > 0$ and the minus/plus sign refers to band index $n=1,2$ respectively. The one-site energy of the bound states, denoted $\epsilon_J(V_0,w)$, and the nearest-neighbor hopping $t_J(V_0,w)$ depend nontrivially on the superlattice potential height $V_0$ and width of the trenches $w$. Within this approximation, the bandwidths are $W_1=W_2=3t_J$ and the bonding--antibonding splitting at $\Gamma$ is $\Delta_{\mathrm{BS}}=E_2(\Gamma)-E_1(\Gamma)=6t_J$. Fig.~\ref{fig:antidot_junction_splitting} plots $\Delta_{\mathrm{BS}}$ obtained directly from the two lowest $\Gamma$-point eigenvalues of Eqs.~\eqref{eq:antidot_hamiltonian} and \eqref{eq:Kronig_Penney_V}, calculated by plane-wave expansion at each $(V_0,w)$. We separately test its interpretation in terms of an effective hopping by fitting the momentum-dependent splitting to Eq.~\eqref{eq:junction_two_orbital} at representative parameter points (Appendix~\ref{app:antidot_numerics}). Over
the sampled momenta, Eq.~\eqref{eq:junction_two_orbital} reproduces the momentum dependence of the splitting with an rms deviation between $0.2\%$ and $4.2\%$ of $\Delta_{\mathrm{BS}}$. Thus
$\Delta_{\mathrm{BS}}/6$ gives an operational estimate of $t_J$ where the nearest-neighbor approximation is accurate. Longer-range couplings or staggered potential (sublattice dependent $\epsilon_J$) would give
corrections. The triangular-junction model introduced below provides a separate approximate prediction of $t_J(V_0,w)$, compared with the continuum results in Fig.~\ref{fig:triangle-junction}.

The bandwidth of each {\it graphene-like} band, $W_1=W_2=\Delta_{\mathrm{BS}}/2$, is
nonmonotonic in $w$ (Fig.~\ref{fig:antidot_junction_splitting}). At large $w$, neighboring junction orbitals overlap efficiently through the wide potential-free regions. Narrowing the trenches initially weakens this overlap. When the width $w$ is further reduced to small enough values for a fixed barrier height, the confinement energy of the shrinking junction region rises and gets closer to $V_0$. Then the wave functions penetrate more strongly into neighboring hexagonal cells. Between the two limits of wide and narrow trenches, we observe a broad minimum in $\Delta_{\mathrm{BS}}$. 

To understand this coupling in a conceptual way, we consider the wave function in an infinitely long trench and, separately, those on a pair of neighboring junctions.
The \emph{transverse trench modes} describe the wavefunction profiles across the trench, perpendicular to its length. A mode confined across the trench may still propagate or decay along it, depending on its total energy. The lowest transverse energy of the trench, $\mu_1$, sets the minimum energy for propagation along it.
For a junction state below $\mu_1$, the wavefunction instead decays exponentially along the trench away from the junction, giving rise to an \emph{evanescent mode}.
Overlap of these tails from neighboring junctions produces a bonding-antibonding splitting and hence the effective hopping $t_J$.
The decay rate is set by the energy difference from $\mu_1$, while the coupling strength also depends on the junction geometry. For finite $V_0$, this threshold depends on both $w$ and $V_0$. In the hard-wall limit $V_0\to\infty$, it approaches $\hbar^2\pi^2/(2mw^2)$; the finite-barrier calculation is given in Appendix~\ref{app:antidot_numerics}.

We estimate the parameters $\epsilon_J$ and $t_J$ using a simple geometrical model of the $Y$ junction connected to straight trenches. Geometrically, the triangular approximation arises very naturally in the narrow-trench limit. Indeed, if the three conducting trenches meeting at a $Y$ junction are idealized as straight rectangular channels, the potential-free region left between them is an equilateral triangle. Remarkably, this geometry is also analytically convenient, since optical modes of equilateral triangular resonators admit analytical constructions based on the triangular symmetry and boundary conditions~\cite{Chang2000,McCartin2004}. This makes it possible to construct a simple semi-analytical model of the junction-centered states while retaining the actual local geometry of the antidot network.  In the present geometry the triangle side has a length $w$ and the neighbouring triangles are connected with the trench of a length $h$.

The lowest fully symmetric state inside the triangular core admits an analytical solution. We approximate the coupling to each trench by a scalar, energy-dependent boundary condition $\partial_n \Psi + \beta(E) \Psi = 0$, where $\beta(E)$ is the logarithmic derivative imposed by the longitudinal wave function in the adjoining trench. For $E<\mu_1$, a finite link connecting two neighboring junctions gives the bonding and antibonding coefficients
\begin{equation}
\begin{aligned}
    \beta_+(E) &=
    \gamma(E)
    \tanh\!\left[\frac{\gamma(E)h}{2}\right], \\
    \qquad
    \beta_-(E) &=
    \gamma(E)
    \coth\!\left[\frac{\gamma(E)h}{2}\right],
\end{aligned}
\end{equation}
where $\gamma(E)=\sqrt{2m(\mu_1-E)}/\hbar$. Depending on $V_0$ and $w$, the antibonding energy can exceed $\mu_1$; the corresponding propagating-link expressions are given in Appendix~\ref{app:triangle_junction}.

Solving the triangular secular equation with $\beta_+$ and $\beta_-$ gives the bonding and antibonding energies $E_b$ and $E_{ab}$. The effective two-orbital parameters then follow directly as

\begin{equation}
    \epsilon_J^\triangle = \frac{E_b + E_{ab}}{2},
    \qquad
    |t_J^\triangle| = \frac{E_{ab} - E_b}{6}.
\end{equation}

This is how both junction energy and hopping parameter could be deduced from the local confinement problem.

\subsection{\textit{Kagome-like} quasi flat band $E_3(\vk)$ \label{sec:2DEGantidotsTrenchStates}}

Band $E_3(\vk)$ is the lowest nearly flat member of the {\it kagome-like} spectrum. In the narrow-band regime, its width $W_3$ is several orders of magnitude smaller than the splitting $\Delta_{\mathrm{BS}}$ between the {\it graphene-like} bands. It touches its dispersive partner $n=4$ at $\Gamma$ and is separated by a finite gap from the {\it graphene-like} bands ($n=1,2$) discussed in Sec.~\ref{sec:2DEGantidotsBoundStates}.

\begin{figure}[t]
    \centering
    \includegraphics[width=\columnwidth]{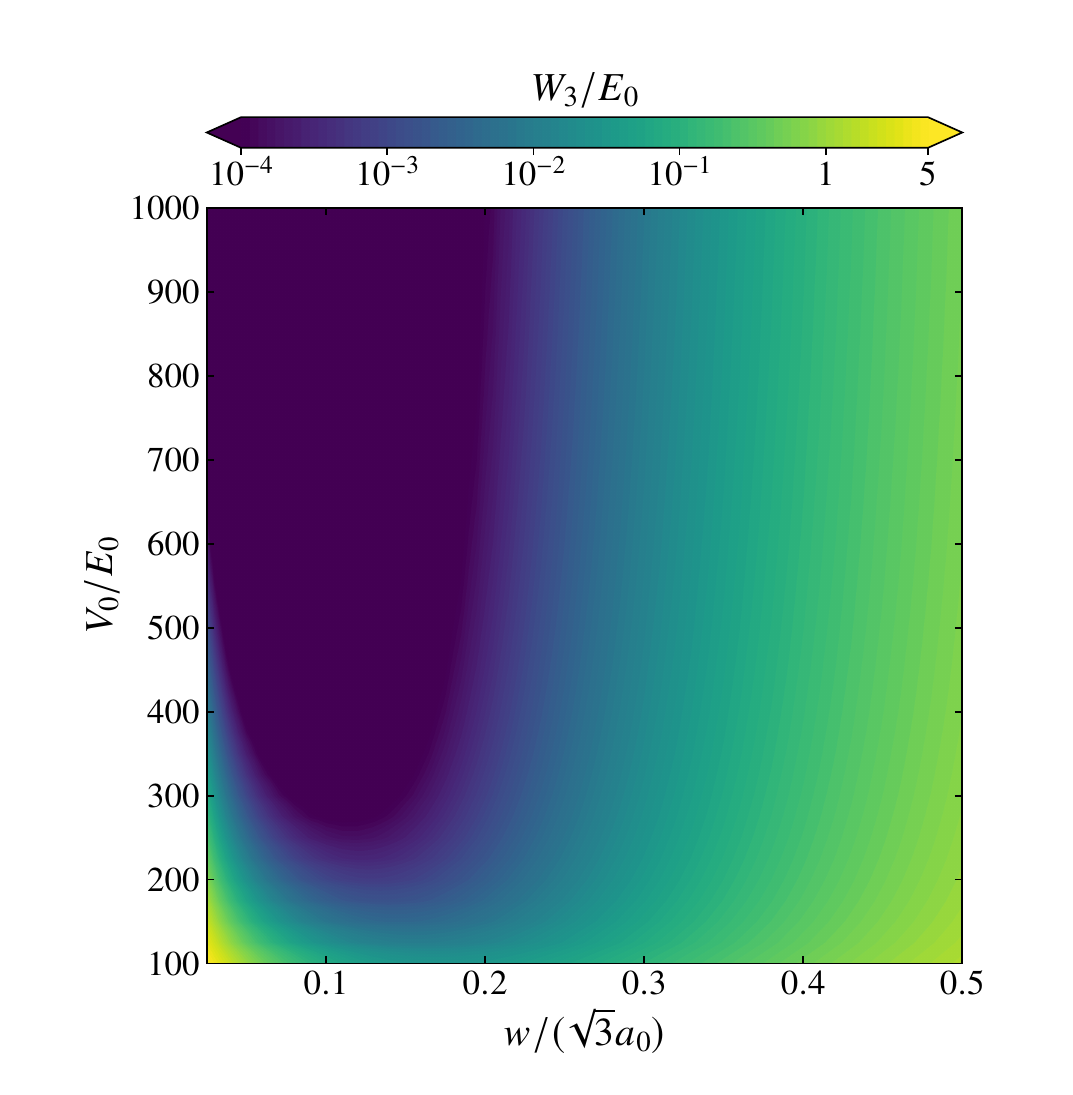}
    \caption{Bandwidth $W_3$ of band $n=3$, the lowest nearly flat member of the {\it kagome-like} spectrum,
    as a function of antidot strength $V_0$ and trench width $w$.
    The logarithmic color scale covers
    $10^{-4}\leq W_3/E_0\leq5$; values below $10^{-4}$ share the lowest
    color.  Values below $5\times10^{-3}E_0$ delineate the narrow-band valley
    but are not interpreted as relatively converged microscopic widths.}
    \label{fig:antidot_bandwidth_map}
\end{figure}
Fig.~\ref{fig:antidot_bandwidth_map} shows the dependence of $W_3$
on $w$ and $V_0$, calculated from the periodic continuum spectrum as
described in Appendix~\ref{app:antidot_numerics}.
Wide trenches couple neighboring loop-like trench states and give
a large bandwidth $W_3$. Increasing the barrier height $V_0$ and narrowing the trenches suppresses
this coupling, producing the broad domain/valley in parameter plane $(w,V_0)$ where this band is nearly flat.
We now turn to the possible mechanisms yielding this dispersion. Anticipating the next sections, a one-dimensional single channel quantum graph picture would provide an exactly flat band. Hence the deviation from flatness is likely to arise from the extension of wavefunctions outside the guiding trench and junction regions. 

Two processes can contribute to the residual dispersion of band $n=3$. The first process concerns \emph{transverse-mode conversion:} in the single-propagating-mode regime of the third band, the lowest transverse trench mode propagates along the links. Evanescent modes therefore play a different role here from the decaying junction-state tails of {\it graphene-like} bound states discussed in Sec.~\ref{sec:2DEGantidotsBoundStates}. Scattering at a junction can excite a higher transverse component whose threshold lies above the third-band energy.
This component decays along the trench, but can reach a neighboring junction and convert back into the propagating mode. Such mode conversion modifies the single-channel interference and can give the third band a finite width $W_3$.
The leading discrete contribution comes from the first odd transverse mode, provided it remains confined across the trench and
is evanescent along it at the energy of the third band.

The second process involves \emph{through-antidot tunneling}: a finite antidot does not perfectly isolate the trench segments that
meet different sides of the same barrier.  Wave-function penetration through
the antidot produces additional nonlocal transfer paths.  Both effects vanish
in the ideal limit of impenetrable barriers and strictly one-dimensional
links.  They act as signed matrix corrections in the subspace containing the
narrow band and the dispersive band that touches it at $\Gamma$; their
positive bandwidth scales therefore cannot, in general, be added as
independent scalar contributions.

\subsection{Conclusion}
We have studied a 2DEG patterned by a triangular array of repulsive
hexagonal antidots. Under strong confinement, its spectrum develops two
{\it graphene-like} bands and higher {\it kagome-like} triplets containing nearly flat
bands. The {\it graphene-like} bands arise from hybridization of junction-centered
orbitals, while the {\it kagome-like} states have weight along the trenches.
Their energy scales respond differently to $V_0$ and $w$: the splitting
$\Delta_{\mathrm{BS}}$ has a broad minimum as a function of $w$, whereas
the width $W_3$ can be orders of magnitude smaller
[Figs.~\ref{fig:antidot_junction_splitting} and
\ref{fig:antidot_bandwidth_map}].

Transverse-mode conversion and through-antidot tunneling can give
the nearly flat {\it kagome-like} bands a finite residual width. For high barriers
and narrow trenches where only the lowest transverse mode propagates,
a single-channel honeycomb quantum graph provides an effective description
of these bands. We develop this limiting model and derive the dashed
spectrum of Fig.~\ref{fig:antidot_graph_crossover}(c) in the next section.

\section{Single channel honeycomb quantum graph\label{sec:singlechannel}} 

In this section, we investigate the single-channel honeycomb quantum graph (HQG) shown in Fig.~\ref{fig:conventions}. For the {\it kagome-like} bands of the antidot model
(Sec.~\ref{sec:2DEGantidots}), only the lowest transverse trench mode is retained
in the regime where it is the only propagating one. Graph bonds describe
its motion along the trenches, while an effective scattering matrix
$S_v(E)$ relates incoming and outgoing amplitudes at each two-dimensional
Y junction. This reduction neglects evanescent propagation between
junctions and through-antidot tunneling, the corrections discussed in
Sec.~\ref{sec:2DEGantidotsTrenchStates}.

The junction matrix generally depends on energy. When its
reflection and transmission amplitudes vary little across the energy
window of interest, we may approximate it by the constant matrix
$S_v(E_\ast)$ at a representative energy $E_\ast$. The general treatment
below also allows to include energy-dependent scattering.

Within this single-channel description, we show how $D_3$-symmetric
junction scattering and destructive interference produce exact flat bands,
extending earlier quantum-graph results
\cite{Kuchment2007,Pal2019,Deoliveira2022}. We construct their compact
localized states and examine the accompanying dispersive bands. A model
with delta-function vertex potentials \cite{Exner1995} then provides a
concrete example and a comparison with the antidot model. The same
framework is relevant to other network realizations
\cite{Efimkin2018,Pal2019,Debeule2023,Zandvliet2025}.

\subsection{Graph model}

\begin{figure}[h!]
\begin{center}
\includegraphics[width=.49\textwidth]{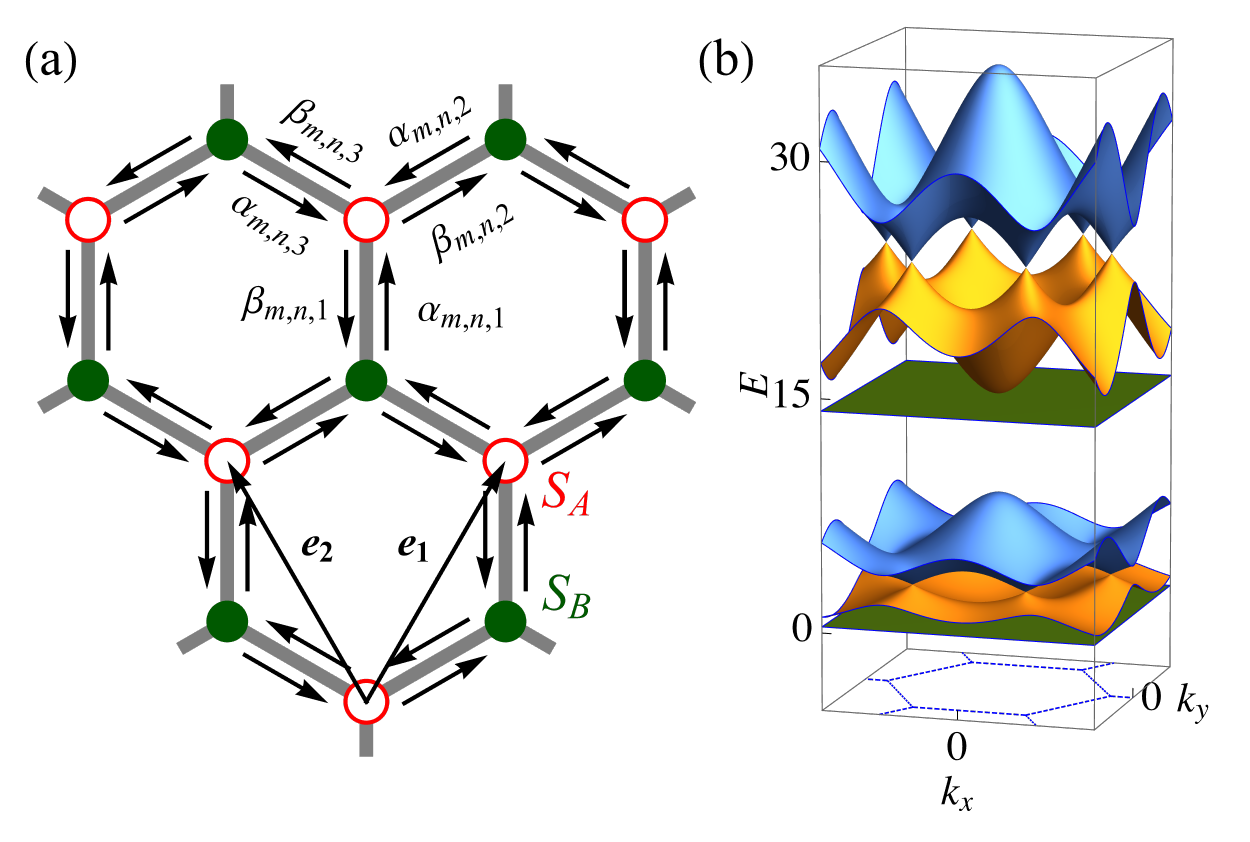}
\caption{(a) The single channel honeycomb wire network, with translation vectors $\ve_{1,2}$ defined by Eq.~\eqref{defe1e2}. Each bond is a conducting wire guiding two counter-propagating electronic plane waves with amplitude $\alpha_{m,n,i}$ and $\beta_{m,n,i}$, $i=1,2,3$ being the leg/wire index and $(m,n)$ labelling the unit cell. Elastic scattering at the nodes is described by the $D_3$-symmetric unitary matrices $S_A$ for sublattice $A$ (open red) and $S_B$ for sublattice $B$ (filled green). (b) Energy spectrum as a function of the Bloch momentum $\vk$ in the Brillouin zone (dashed blue hexagon). The plotted coordinates are $E/E_0$ and $\vk a_0$. Flat bands (green) occur for arbitrary scattering matrices obeying the $D_3$ symmetry. Here we have chosen energy-independent scattering matrices Eq.~\eqref{Ssinglebis} parametrized by $(\theta_v,\phi_v)=(\pi/9,4\pi/5)$ for both $v=A,B$.}\label{fig:conventions} 
\end{center}
\end{figure}

In the quantum graph model, the Schrödinger differential operator acts along the 1D bonds of the honeycomb graph and reads $H = -\hbar^2/(2m)d^2/ds^2$, where $s \in (0,a_0)$ denotes a continuous local coordinate defined on each bond, as the distance from the $B$-site ending this bond. The bonds of the graph can be seen as the conducting channels/trenches of the 2D antidot model in the limit $w \rightarrow 0$, or alternatively as edge states surrounding the hexagonal voids of the honeycomb structure \cite{Efimkin2018,Pal2019}. We consider solely elastic scattering at the nodes described by 3 by 3 scattering matrices $S_v$, indexed by the sublattice index $v=A,B$. The bond index $\vl = (m,n,i)$ encapsulates the leg index $i \in \{1,2,3\}$, and the integer coordinates $(m,n) \in \mathbb{Z}^2$ for the unit cell (Fig.~\ref{fig:conventions}). The scalar wave function $\Psi_\vl(s)$ at energy $E$ inside bond $\vl = (m,n,i)$ is a linear superposition of two counterpropagating plane waves : 
\begin{equation}\label{Psi}
    \Psi_\vl(s)= \alpha_\vl \,\, e^{iq s}  + \beta_\vl \,\, e^{-iq s} \, \, ,
\end{equation}
where $q>0$ is the longitudinal wave vector along the bond. The energy is $E=\hbar^2q^2/(2m)$, or equivalently in dimensionless form $E/E_0=(qa_0)^2$.

At a node on the $A$-sublattice ($s=a_0$), the matching conditions read : 

\begin{equation}\label{eq:matchingrealspace1}
\begin{pmatrix}
e^{-iqa_0}\beta_{m,n,1}\\
e^{-iqa_0}\beta_{m,n,2}\\
e^{-iqa_0}\beta_{m,n,3}
\end{pmatrix}
=
S_A
\begin{pmatrix}
e^{iqa_0}\alpha_{m,n,1}\\
e^{iqa_0}\alpha_{m,n,2}\\
e^{iqa_0}\alpha_{m,n,3}
\end{pmatrix},
\end{equation}
while at a $B$-type site ($s=0$) :
\begin{equation}\label{eq:matchingrealspace2}
\begin{pmatrix}
\alpha_{m,n,1}\\
\alpha_{m-1,n,2}\\
\alpha_{m,n-1,3}
\end{pmatrix}
=
S_B
\begin{pmatrix}
\beta_{m,n,1}\\
\beta_{m-1,n,2}\\
\beta_{m,n-1,3}
\end{pmatrix}.
\end{equation}
Owing to the local $D_3$ symmetry of each HQG vertex, the scattering matrix commutes with the permutation representation of $D_3$ on the three legs and has the structure
\begin{equation}\label{eq:Srt}
S_v=
\begin{pmatrix}
r_v&  t_v & t_v \\
t_v&  r_v & t_v \\
t_v&  t_v & r_v 
\end{pmatrix} \, \, ,
\end{equation}
where the complex numbers $r_v$ and $t_v$ are respectively the reflection and transmission amplitudes of type $v=A,B$ vertex. Their energy dependence, like that of $S_v$, is left implicit below \footnote{Approximating the junction matrix as constant can fail near junction resonances or transverse-mode thresholds. Moreover, a constant effective matrix need not correspond to an energy-independent self-adjoint boundary condition at a point vertex~\cite{Kostrykin1999}.}. The comparison with the antidot bands in Fig.~\ref{fig:antidot_graph_crossover}(c) retains this energy dependence using the junction calculation of Appendix~\ref{app:antidot_numerics} and Sec.~\ref{subsec:deltascatterers} gives a delta-function scattering model with an explicit energy dependent $S_v(E)$. The spectral decomposition reads
\begin{equation}\label{Ssinglebis}
   S_v =e^{i \phi_v}  P_\perp + e^{i \theta_v } P_\parallel  \,\, ,
\end{equation}
where the eigenvalue $e^{i \theta_v }$ is associated with the rank-1 projector $P_\parallel$ along the vector $\ket{e_3} = (1,1,1)/\sqrt{3}$, while eigenvalue $e^{i \phi_v }$ is attached to the rank-2 projector $P_\perp = \mathbb{I}_{3}- P_\parallel$ onto the subspace orthogonal to $\ket{e_3}$. The scattering matrix $S_v$ can be represented either by Eq.~\eqref{eq:Srt} or parametrized by the angles $(\theta_v,\phi_v)$ according to Eq.~\eqref{Ssinglebis}, the correspondence being given by $t_v = (e^{i\theta_v} - e^{i\phi_v})/3$ and $r_v =(e^{i\theta_v} + 2 e^{i\phi_v})/3 $.

Taking advantage of the periodicity of the honeycomb lattice, the Bloch theorem \begin{equation}
 \Psi_{(m,n,i)}(s) = e^{i \vk \cdot(m\ve_1+n\ve_2)} \Psi_{(0,0,i)}(s) \, \, ,  
 \end{equation}
can be used, $\vk$ being the Bloch lattice wave vector. The Brillouin zone is the one of the 2D honeycomb lattice with high-symmetry points $\Gamma$, $M$ and $K$, as in Sec.~\ref{sec:2DEGantidots}. Under the Bloch reduction $(m,n) \rightarrow \vk$, the scattering conditions at vertices $A$ and $B$ read :
\begin{equation}\label{eq:matching}
\begin{pmatrix}
|\alpha_\vk\rangle \\
|\beta_\vk \rangle
\end{pmatrix}=
\begin{pmatrix}
0&  D^\dagger_\vk S_B D_\vk\\
e^{2iq a_0 } S_A & 0
\end{pmatrix}
\begin{pmatrix}
|\alpha_\vk\rangle \\
|\beta_\vk \rangle
\end{pmatrix} \, \, ,
\end{equation}
where the vector $\ket{\alpha_\vk} =(\alpha_{\vk,1},\alpha_{\vk,2},\alpha_{\vk,3})$ collects the amplitudes of modes propagating from $B$ to $A$ along the 3 different legs, while $\ket{\beta_\vk} =(\beta_{\vk,1},\beta_{\vk,2},\beta_{\vk,3})$ represents the amplitudes of the modes propagating from $A$ to $B$. The diagonal matrix $D_\vk$ with diagonal entries $(1,e^{-i\vk\cdot \ve_1},e^{-i\vk \cdot \ve_2})$ accounts for the Bloch phase shifts between different unit cells (Fig.~\ref{fig:conventions}). 

The condition for a non-trivial solution of Eq. \eqref{eq:matching} is expressed by the following eigenproblem :
\begin{equation}\label{seculareqtrois}
 e^{2 iq(E) a_0} D_\vk^\dagger S_B D_\vk S_A \ket{\alpha_\vk} = \ket{\alpha_\vk}   \, ,
\end{equation}
where $q=q(E)=\sqrt{2mE}/\hbar$. Solving the corresponding determinant secular equation 
\begin{equation}\label{detseculareqtrois}
 \det \left[e^{2 iq(E) a_0} D_\vk^\dagger S_B D_\vk S_A -\mathbb{I}_{3} \right]=0
\end{equation}
provides the full band structure of HQG, which consists of 3 branches of $e^{2iq a_0}$ vs $\vk$, yielding an infinite spectrum of discrete levels $q$ (or $E$) for each $\vk$. Note that this quantum graph construction, and specifically Eq.~\eqref{seculareqtrois}, apply to any bipartite lattice and generically leads to dispersive bands. For the particular case of the HQG, it is shown in the next subsection that for each Bloch wave vector $\vk$, one can find a nontrivial vector $\ket{\alpha_\vk}$ that solves the eigenproblem Eq.~\eqref{seculareqtrois} with a value of $q$ \emph{independent} of $\vk$. This flat band solution is not guaranteed to exist for arbitrary bipartite lattice because the scattering matrices do not commute with $D_\vk$. 

\subsection{Flat bands}
Now we establish a direct link between the structure of the scattering matrix Eq. (\ref{Ssinglebis}) and the existence of robust symmetry protected FB for the singlechannel HQG. We explicitly construct a Bloch state that solves Eq.~\eqref{seculareqtrois}. 
For any $\vk$, one chooses a Bloch vector $\ket{\alpha_{\vk}}$ such that 
\begin{equation}
    \ket{\alpha_\vk} \in \text{Im} P_\perp \hspace{5mm} {\rm and} \hspace{5mm} D_\vk \ket{\alpha_\vk} \in \text{Im} P_\perp \, \, .
\end{equation}
Then for such states, the scattering matrix acts simply as $S_A \ket{\alpha_{\vk}} =e^{i\phi_A} \, \ket{\alpha_{\vk}}$ and $S_B D_\vk \ket{\alpha_{\vk}} =e^{i\phi_B} D_\vk \ket{\alpha_{\vk}}$. Therefore the eigenproblem Eq.~\eqref{seculareqtrois} reduces to :
\begin{equation}
 e^{i(2 q \,a_0 +  \phi_A +  \phi_B)}  \ket{\alpha_\vk} = \ket{\alpha_\vk}   \, 
\end{equation}
since $D_\vk^\dagger D_\vk = \mathbb{I}_{3}$. The corresponding solutions $q$ obey 
\begin{equation}\label{flatband}
 2 q \,a_0  +\phi_A + \phi_B =  2 \pi p  \,\, ,
\end{equation}
where $p$ is an integer. The bond momentum $q$ and the energy $E=\hbar^2q^2/(2m)$ are $\vk$-independent, thereby
giving a flat band over the whole Brillouin zone for each allowed value of $p$. The corresponding scattering eigenvector forms the nontrivial texture in Bloch momentum 
\begin{equation}\label{eq:alphak_singlechannel}
\vk  \mapsto \ket{\alpha_\vk}=\frac{1}{\sqrt{N_\vk}}
\begin{pmatrix}
b_\vk-a_\vk\\
1-b_\vk\\
a_\vk-1
\end{pmatrix}\,
\end{equation}
where $a_\vk=e^{-i\vk\cdot\ve_1}$, $b_\vk=e^{-i\vk\cdot\ve_2}$ and :
\begin{equation}
N_\vk
= |b_\vk -a_\vk|^2+|1-b_\vk|^2 + |a_\vk-1|^2 \, .
\end{equation}
This normalized ket is locally the direction orthogonal to the plane spanned by $\ket{e_3} = (1,1,1)/\sqrt{3}$ and $D_\vk^\dagger\ket{e_3}$. Note that at $\vk=0$, this direction is ill-defined which leads to a singular band touching between each FB and a dispersive band \cite{Bergman2008,Rhim2019}. 

Hence, even for arbitrary values of $\theta_v$ and $\phi_v$, Eq.~\eqref{flatband} coincides with the Bohr–Sommerfeld quantization rule one would have for the very particular case of perfect mirrors with zero transmission $t_v=0$ and unimodular total reflection $r_v$, namely when $\theta_v=\phi_v$ and $S_v = e^{i \phi_v} \mathbb{I}_3$, since $2qa_0+\phi_A+\phi_B$ is the phase accumulated along a closed semi-classical trajectory. 

Note that the energy of the FBs depends only on the angular parameters $\phi_v$ ($v=A,B$), and not on $\theta_v$, which affect solely the dispersive bands. Tuning the angle $\theta_v$ allows to further flatten the dispersive bands. 

Previous works \cite{Kuchment2007} treated the particular case of the Kirchhoff--Neumann boundary conditions, where the wave function is continuous and the sum of its outward derivatives vanishes at each vertex, which correspond to the very specific parameters $(\theta_v,\phi_v)=(0,\pi)$ for $v=A,B$. The linear algebraic argument above shows that the FB are robust for any matching conditions provided they respect unitarity and $D_3$ symmetry between the three legs at each Y junction.  

Finally, our analysis also allows for different scattering matrices, $S_A$ and $S_B$, at $A$-type and $B$-type vertices, which generalizes previous work assuming $S_A=S_B$  \cite{Pal2019} to staggered honeycomb lattices. The robust FBs of the HQG are also related to the FBs predicted in the tight binding model for kagome ($S_A=S_B$) and breathing kagome \cite{Ezawa2018,Bolens2019} ($S_A \neq S_B$) lattices.

\subsection{Compact localized states} 
The extended Bloch states within a given FB are all degenerate in energy and it is possible to construct superpositions of such Bloch states that have compact supports \cite{Bergman2008,Rhim2019,Hwang2021,Graf2021}. For the HQG, the smallest compact localized state (CLS) corresponds to a wave function vanishing everywhere on the network but on a single hexagon. 
\begin{figure}[h]
\centering
\includegraphics[width=0.3\textwidth]{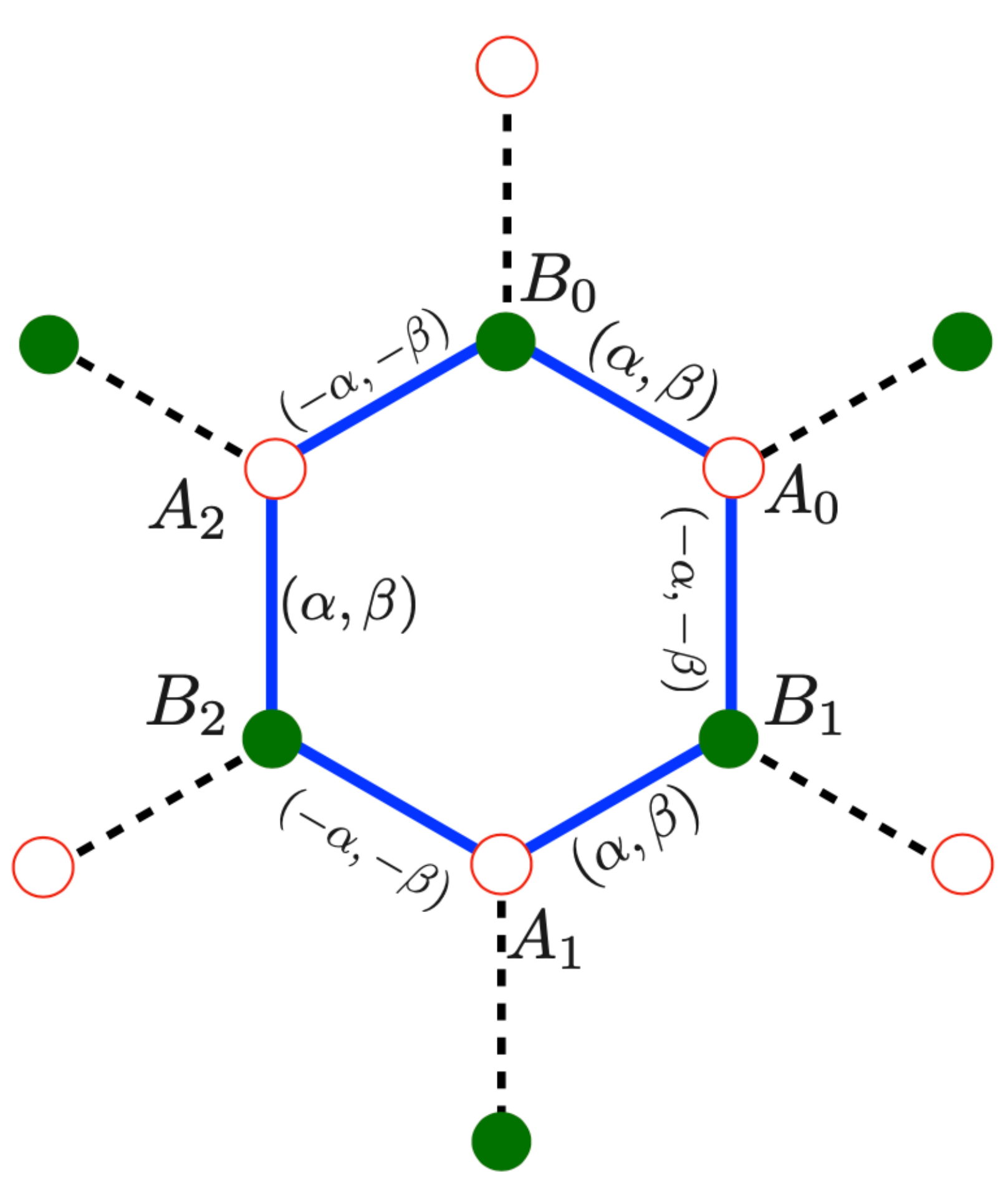}
    \caption{Compact localized state construction. The wave function is non zero along the (blue) solid line hexagon $\mathcal{H}$ and vanishes everywhere else on the graph, in particular along the dashed bonds.}\label{figCLS}
\end{figure}
Here we build explicitly such a CLS corresponding to the FBs Eq.~(\ref{flatband}) of the HQG, using the real space framework. The construction consists in three steps.

First, we select a hexagon $\mathcal{H}$ in the HQG with corners $B_0 A_0 B_1 A_1 B_2 A_2$ (Fig.~\ref{figCLS}). We set all amplitudes $\alpha_{m,n,i}$ and $\beta_{m,n,i}$ to zero except on $\mathcal{H}$. Second, we choose the scattering states on $\mathcal{H}$ to follow an alternating pattern with amplitudes
\begin{equation} 
(\alpha,\beta) (-1)^{a-b}  \,\, ,
\end{equation}
on the bond $A_a B_b$ (Fig.~\ref{figCLS}). The complex numbers $\alpha$ and $\beta$ are arbitrary for the moment. This construction leads to scattering in- or out- states that are orthogonal to the vector $\ket{e_3}=(1,1,1)/\sqrt{3}$. Therefore Eq. (\ref{eq:matchingrealspace1}) at $A_0$ reads :  
\begin{equation}\label{eq:matchingrealspaceA0}
e^{-2iqa_0}
\begin{pmatrix}
-\beta\\
0\\
\beta
\end{pmatrix}
=
S_A
\begin{pmatrix}
-\alpha\\
0\\
\alpha
\end{pmatrix}
=
e^{i \phi_A}
\begin{pmatrix}
-\alpha\\
0\\
\alpha
\end{pmatrix}
,
\end{equation}
which reduces simply to $\beta=\alpha \, e^{i(2qa_0+\phi_A)}$. Similarly, Eq.~\eqref{eq:matchingrealspace2} at $B_0$ reads :  
\begin{equation}\label{eq:matchingrealspaceB0}
\begin{pmatrix}
0\\
-\alpha\\
\alpha
\end{pmatrix}
=
S_B
\begin{pmatrix}
0\\
-\beta\\
\beta
\end{pmatrix}
=
e^{i \phi_B}
\begin{pmatrix}
0\\
-\beta\\
\beta
\end{pmatrix}
,
\end{equation}
which reduces to $\alpha = \beta \, e^{i \phi_B} $. Third, gathering Eqs.~\eqref{eq:matchingrealspaceA0} and \eqref{eq:matchingrealspaceB0} leads to the condition :  
\begin{equation}
 e^{i(2 q a_0 +  \phi_A +  \phi_B)}  = 1   \, 
\end{equation}
to allow a non-trivial state ($\alpha \neq 0$), which is the real-space version of the Bohr-Sommerfeld condition Eq. (\ref{flatband}) already obtained in the Bloch representation. This construction can be performed for any hexagon on the graph which explains the macroscopic degeneracy of the FBs. It is possible to go from CLS to Bloch scattering states by summing the CLS over unit cells with appropriate phase, and vice versa.

\begin{figure}
\centering
\includegraphics[width=0.49\textwidth]{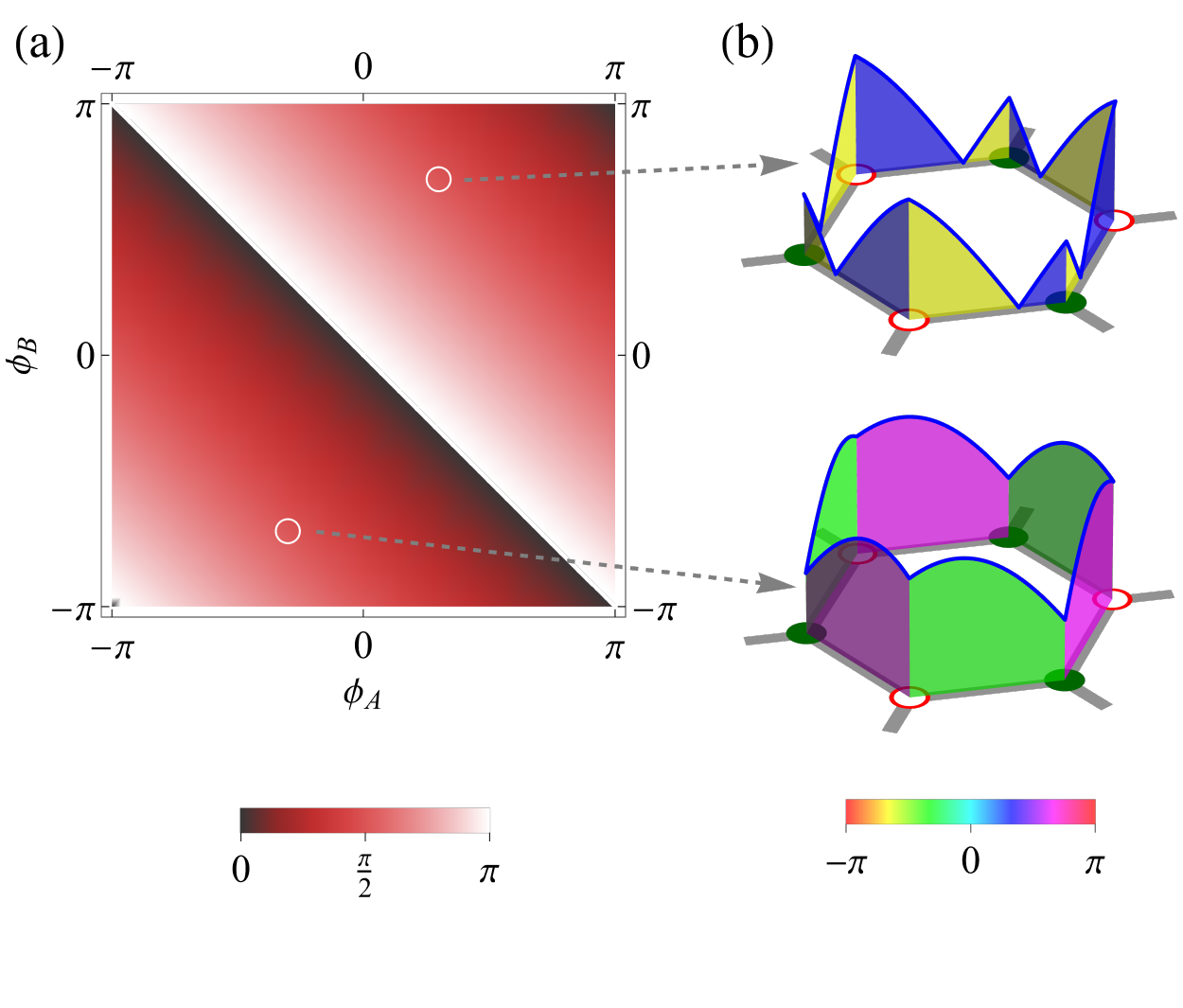}
\vspace{-1.5cm}
    \caption{Lowest-energy compact localized states (CLSs). (a) The dimensionless wave number $qa_0$ (color scale) as a function of $\phi_{A,B}$, with $qa_0/\pi\equiv-(\phi_A+\phi_B)/(2\pi)~\mathrm{mod}~1$. (b) Two representative CLSs for $(\phi_A,\phi_B)=\pm(3\pi/10,7\pi/10)$; height and color show respectively the amplitude and phase of the wave function. A CLS in the lower-left (upper-right) triangular region of panel (a) has zero (one) internal node, as shown in panel (b).}\label{fig2}
\end{figure}

For the particular case of Neumann vertex matching, $\phi_A=\phi_B=\pi$, we recover the Dirichlet eigenstates of Ref.~\cite{Kuchment2007}, for which the CLSs have nodes at the vertices. The same interference construction remains valid in the presence of an identical longitudinal potential $V(s)$ on every bond, provided the bond orientation and vertex symmetries are preserved. This extends the setting of Ref.~\cite{Kuchment2007}, where $V(s)$ was assumed symmetric about the center of a bond.

More generally, the number and locations of the nodal points in the CLS, depend on the FB number $n$ and on the parameters $(\phi_A,\phi_B)$. The energies of the various FBs are obtained by the formula
\begin{equation}
 2 q \, a_0  +\phi_A + \phi_B =  2 \pi p  \, , \hspace*{5mm} q>0  \, ,\
\end{equation}
where $p$ is an integer which leads to the quantization of $q=q_p>0$, and can be used as a {\it flat band index}. The energy is given by $E_p=\hbar^2q_p^2/(2m)$ and the wave function Eq.~(\ref{Psi}) reads :
\begin{equation}
    \Psi_p(s)=2 \alpha e^{-i\phi_B /2} \cos \left( q_p \, s + \frac{\phi_B}{2}\right)  \, \, ,
\end{equation}
since $\alpha e^{-i\phi_B /2}=\beta e^{i\phi_B /2}$. The condition for a node at the $B$ site is $\Psi(0)=0$, hence $\phi_B=\pm\pi$. Similarly, a node at the $A$ site requires $\Psi(a_0)=0$, hence $\phi_A=\pm\pi$. Thus Neumann vertex matching gives CLSs with Dirichlet nodes at both endpoints.

Now let us investigate the condition for observing a node at local coordinate $0<s<a_0$ strictly located at the interior of a bond : 
\begin{equation}
     q_p \, s + \frac{\phi_B}{2} = \frac{\pi}{2} + l \pi
\end{equation}
Since $q_p>0$ and $0<s<a_0$, we have :
\begin{equation}
    0<  q_p \, s  < q_p \, a_0 = \pi p -  \frac{\phi_A+\phi_B}{2} 
\end{equation}
The integer $l$ has therefore to satisfy the inequalities :
\begin{equation}
 \frac{\phi_B}{2} - \frac{\pi}{2}    <\pi l < \pi (p - \frac{1}{2}) - \frac{\phi_A}{2}
\end{equation}
We consider $(\phi_A,\phi_B) \in ]-\pi,\pi[^2 $. The minimal value for $l$ is then
$l=0$, and its maximal value is $l=p-1$. Therefore, the number of nodes along each
bond is equal to $p$, and it increases by one upon going from one flat band to the next one
higher up in energy. Note that when $p=0$, one has a CLS with no nodes at all, a situation which can happen only when $\phi_A + \phi_B < 0$. When $\phi_A + \phi_B > 0$, the minimal
value of $p$ is $p=1$. The difference between these two cases is illustrated on Fig.~\ref{fig2},
where $p$ is assumed to take its minimal value, depending on the sign of $\phi_A + \phi_B$, which corresponds to the lowest energy FB.

\subsection{Dispersive bands \label{sec:dispersivebands}}
Besides the FB discussed above, the HQG spectrum also contains dispersive bands as shown in Fig.~\ref{fig:conventions}(b) and whose exact dispersion relation are detailed in Appendix~\ref{app:singlechannel_dispersive}. There are two generically dispersive bands for a flat band. In the generic case, there are some gaps between the different bands. Nevertheless for certain values of $(\theta_A,\theta_B,\phi_A,\phi_B)$, the two
dispersive bands form Dirac points at :
\begin{equation}
 \vK=\left(\frac{2\pi}{3\sqrt3a_0},\frac{2\pi}{3a_0}\right).
\end{equation}
For other fine-tuning ot the parameters, it is also possible to completely flatten the dispersive bands. This is has to be seen as an accidental degeneracy in contrast with the symmetry protected/induced degeneracy of the generic and robust flat bands described above. 

Apart from the exceptional all-flat cases, each flat band has a topologically protected touching with a dispersive band at $\Gamma$ ($\vk=0$) \cite{Bergman2008,Rhim2019}. This band touching is a singular one because in the limit $\vk \rightarrow 0$, the unnormalized numerator in Eq.~\eqref{eq:alphak_singlechannel} vanishes. When $\vk \neq 0$, the limit of the normalized eigenvector $|\alpha_\vk\rangle$ corresponding to the flat band, given by Eq. \ref{eq:alphak_singlechannel} (and also the ket $|\beta_\vk \rangle = e^{2i q a_0} S_A | \alpha_\vk\rangle$) is ill defined when $k \rightarrow 0$ because it depends on the direction along which the origin is approached. 

\subsection{Example of the Dirac delta function nodal scatterers \label{subsec:deltascatterers}}

The preceding analysis of the singlechannel HQG applies for any arbitrary local vertex scattering obeying the $D_3$-symmetry. Up to this point, the scatterer has been specified only through its scattering matrix $S_v(E)$, without assuming any particular microscopic realisation. We now consider one specific realization of such a vertex scatterer: a scalar $\delta$-like potential of strength $\tau$ placed at each node.
For simplicity, we take identical vertex potentials on the two sublattices, $\tau_A=\tau_B=\tau.$ This example is particularly transparent because it turns out that such scattering model modifies only one of the two scattering sectors introduced in Eq.~( \ref{eq:Srt}).

For a $B-$type vertex ($s=0$), the corresponding boundary conditions are
\begin{equation}
    \begin{aligned}
        &\Psi_1(0)=\Psi_2(0)=\Psi_3(0)=\Psi(0),\\
        \qquad &\sum_{j=1}^{3}\partial_{\mathrm{out}}\Psi_j(0) =\tau\, \Psi(0)\, ,
    \end{aligned}
    \label{eq:delta_boundary}
\end{equation}
where the cell indexes $(m,n)$ are implied in $\Psi_j = \Psi_{m,n,j}$, and $\partial_{\mathrm{out}} \Psi_j$ is the outgoing derivative along leg $j=1,2,3$.  The first condition imposes continuity of the wave function at the vertex. The presence of a delta-like scatterer at the vertex implies a jump $\tau \psi(0)$ for the total outgoing derivative. The boundary parameter $\tau$ has units of inverse length; the corresponding delta-potential strength is $\hbar^2\tau/(2m)$. The parameter \(\tau\) encodes the strength of the vertex potential : \(\tau>0\) corresponds to a repulsive vertex potential, and \(\tau<0\) to an attractive one. Note that $\tau=0$ gives the Kirchhoff-Neumann boundary condition \cite{Kuchment2007}.

For an attractive vertex potential, $\tau < 0$, the same boundary condition also support a localized bound state at an isolated three-leg vertex. Taking $\Psi_j(s)=A e^{-\kappa s}$ with $\kappa>0$ on each outgoing bond, Eq~(\ref{eq:delta_boundary}) gives $-3 \kappa = \tau$. So, such bound states exist only for $\tau < 0$, and their common energy is given by $E_b = -[\hbar^2/(2m)]\tau^2/9$.

In the periodic HQG, these vertex-localized states hybridize between the neighbouring nodes and form dispersive bands. These bands provide a quantum-graph analogue of the {\it graphene-like} bands in the continuum antidot model, which likewise arise from hybridization of local junction states (Sec.~\ref{sec:2DEGantidotsBoundStates}).

This bound state belongs to the negative part of the single-vertex spectrum. We now return to the positive-energy scattering states, $E=\hbar^2q^2/(2m)>0$, which determine the propagating spectrum of the HQG.

The scattering matrix reads  
\begin{equation}
    S_A(q)=S_B(q)=- P_{\perp} + \frac{3q-i\tau}{3q+i\tau} \, P_{\parallel}  \, \, \label{eq:delta_projector_form}
\end{equation}
showing the explicit energy dependence $S_A(q=\sqrt{2mE}/\hbar)$. Comparing this expression with the general decomposition Eq. \eqref{Ssinglebis} we identify
\begin{equation}
    e^{i\phi_A} = e^{i\phi_B} = -1,
    \hspace*{3mm}
    e^{i\theta_A(q)} = e^{i\theta_B(q)} = \frac{3q-i\tau}{3q+i\tau} \,\,.
\end{equation}
Thus the vertex potential changes only the symmetric-sector phase $\theta_v(q,\tau)$, while the $P_\perp$ eigenvalue remains $-1$. The Bloch construction still applies, but its secular equation is implicit in $q$, which enters both the propagation and scattering phases.

The diagonal and off-diagonal elements of $S_v$ define the reflection and transmission amplitudes,
\begin{equation}
    r(q,\tau) = -\frac{q+i\tau}{3q+i\tau},
    \quad
    t(q,\tau) = \frac{2q}{3q+i\tau} \, \, .
    \label{eq:reflection_transmission}
\end{equation}
For \(\tau=0\), one recovers the Kirchhoff-Neumann values \(r=-1/3\) and \(t=2/3\). In the limit $|\tau| \rightarrow \infty$, the transmission amplitude decreases, while the reflection probability approaches unity.

\begin{figure}
\centering
\includegraphics[width=0.49\textwidth]{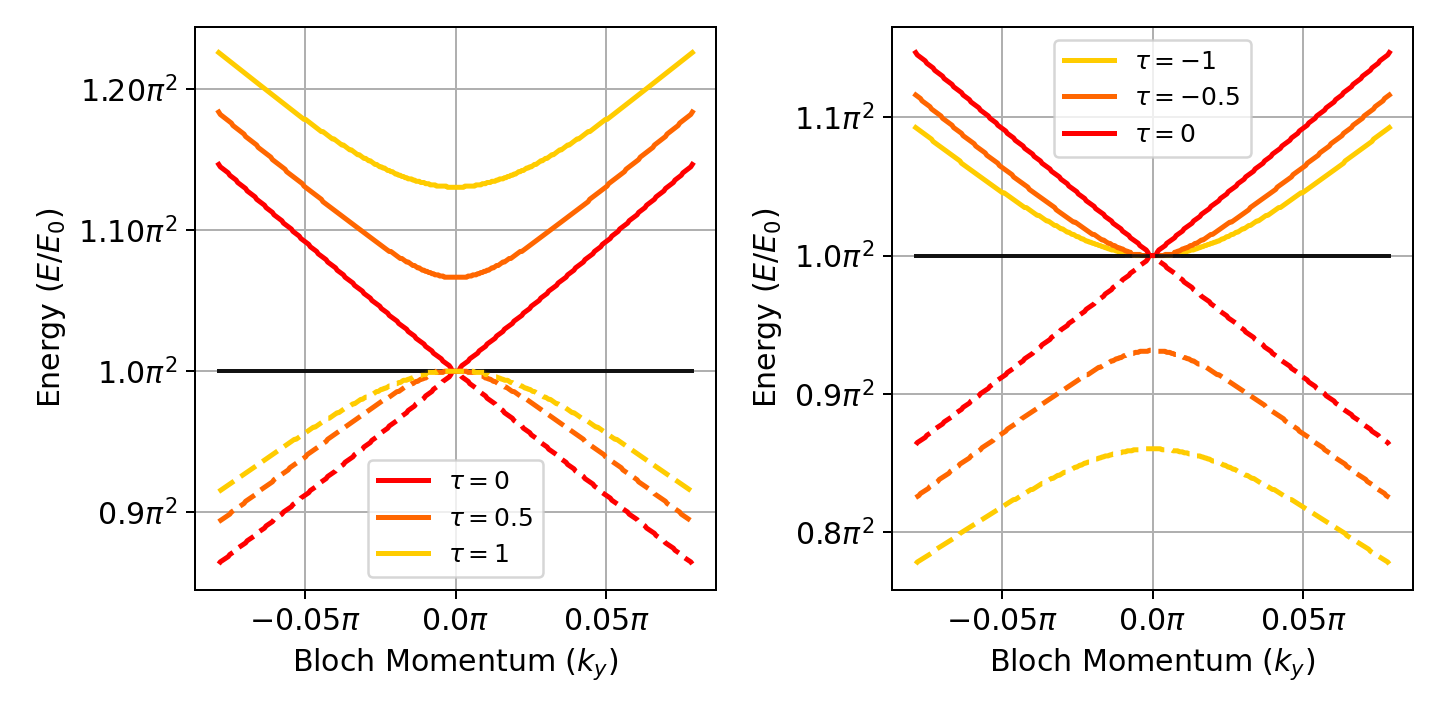}
\vspace{-0.5cm}
\caption{Normal-state spectrum along the cut \(k_x=0\) for positive (left) and negative (right) values of the vertex potential \(\tau\). The horizontal axis and legend values denote $k_y a_0$ and $\tau a_0$, respectively. The horizontal black line represents the pinned flat band, while the solid and dashed curves correspond to the neighboring dispersive bands. The vertex scatterer opens a gap on one side of the flat band, preserves the touching with the opposite dispersive band at \(\Gamma\), and progressively reduces the bandwidth of the dispersive states.}
\label{fig:normal_spectrum_tau}
\end{figure}

The independence of the flat bands from $\tau$ then follows immediately from the general flat-band condition given by Eq.~(\eqref{flatband}). Since $\phi_A=\phi_B=\pi$, one obtains
\begin{equation}
    q_n^{\mathrm{flat}} = \frac{n\pi}{a_0},
    \qquad
    E_n^{\mathrm{flat}} = \frac{\hbar^2}{2m} \, \left( \frac{n\pi}{a_0} \right)^2 \, ,
    \label{eq:flat_band_energies}
\end{equation}
where $n=1,2,\ldots$. The flat-band energies are therefore independent of $\tau$ and correspond to the energies of a nonrelativistic particle in a 1D box with Dirichlet boundary conditions.

The flat bands remain pinned in Fig.~\ref{fig:normal_spectrum_tau} because the corresponding CLSs vanish at every vertex. Their alternating bond amplitudes also make the sum of outward derivatives vanish, so both sides of the $\delta$-matching condition in Eq.~\eqref{eq:delta_boundary} are zero for any $\tau$. The dispersive states generally have nonzero vertex amplitudes and respond to the changing symmetric-sector phase $\theta(q,\tau)$.

A nonzero $\tau$ opens a gap between each flat band and one neighboring dispersive branch. Its sign determines which side of the flat band is gapped. The other branch retains its symmetry-protected touching with the flat band at $\Gamma$, so the flat band is not isolated.

In addition to shifting the dispersive bands, the vertex potential reduces their bandwidth. This flattening can be understood directly from the transmission amplitude in Eq.~\eqref{eq:reflection_transmission}. As \(|\tau|\) increases, \(|t|\) decreases and propagation through the vertices is progressively suppressed. The dependence of the energy on the Bloch momentum consequently becomes weaker, and the dispersive bands become increasingly flat. In the large-\(|\tau|\) limit, the bonds are almost decoupled and the dispersive levels approach the discrete spectrum of isolated one-dimensional segments.

\begin{figure}[t]
    \centering

        \includegraphics[width=\linewidth]{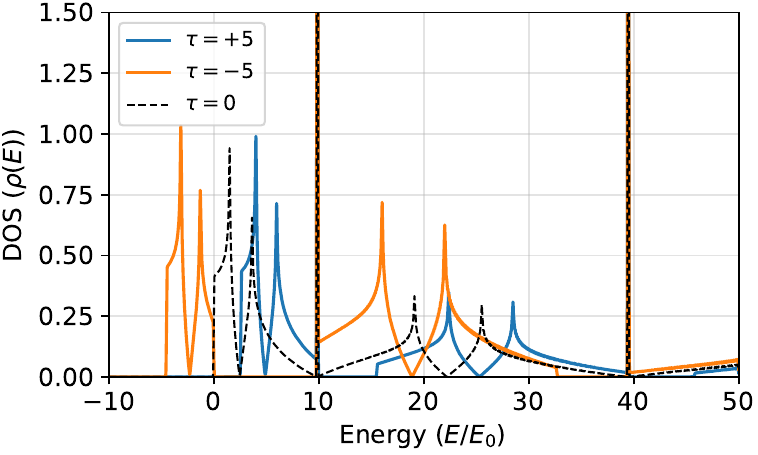}
    \caption{Density of states $\rho(E)$ for attractive and repulsive vertex potential \(\tau\) with same absolute value : $|\tau|a_0=5$; the legend values denote $\tau a_0$. The positions of the flat-band peaks (at $E/E_0 \simeq 10$ and $E/E_0 \simeq 40$) are independent of $\tau$ whereas the DOS structure between the peaks depends on the potential strength : the dispersive spectral weight is shifted. At lowest energies/filling, both spectra show a DOS characteristic of a pair of {\it graphene-like} bands, with two van Hove peaks separated by a Dirac point cancellation of the DOS.}
    \label{fig:dos_tau}
\end{figure}

Fig.~\ref{fig:dos_tau} shows the corresponding density of states, obtained by sampling the Bloch spectrum over the Brillouin zone and accumulating the energies. The flat-band peaks remain fixed, while the broader dispersive contributions shift to opposite sides for positive and negative $\tau$.

\section{Multichannel honeycomb quantum graph
\label{sec:multichannel}}

The single-channel graph model (Sec.~\ref{sec:singlechannel}) isolates the destructive-interference mechanism responsible for the formation of exact flat bands in the HQG due to $D_3$ symmetry around each vertex. With wires of finite width $w$, however, multiple transverse modes become propagating as the energy $E$ exceeds their respective thresholds, introducing additional scattering channels at each vertex. In this section, we investigate the conditions under which the flat-band construction established in Sec.~\ref{sec:singlechannel} survives in the presence of such multimode scattering. We show that mixing between transverse modes of the same parity preserves the flat-band construction, whereas coupling between modes of opposite parity generically lifts the flat-band degeneracy and yields dispersive bands. This distinction originates from the symmetry constraints on the vertex-scattering matrix and the different propagation properties of the two parity sectors between neighboring vertices. This multichannel HQG model can also be seen as intermediate between the patterned 2DEG (Sec. \ref{sec:2DEGantidots}) and the single channel HQG. 

\subsection{Finite-width graph model
\label{sec:multichannel_graph}}

We now consider a fattened graph model, which is basically the HQG of Sec.~\ref{sec:singlechannel} with a finite width $w$. All the links have identical length $a_0$ and width $w$. We attach to each link a system of longitudinal and transverse local coordinates
$(s,s_\perp)$, where $0<s<a_0$ and
$-w/2<s_\perp<w/2$.  For hard lateral walls, we denote $\Phi_j(s_\perp)$ the
normalized transverse eigenfunctions and $\mu_j$ their threshold
energies, arising from the increase in kinetic energy due to lateral confinement. The propagating longitudinal momenta $q_j(E)$ at
energy $E$ are given by
\begin{equation}
 q_j(E)=\frac{\sqrt{2m(E-\mu_j)}}{\hbar},
 \qquad
 \mu_j=\frac{\hbar^2}{2m}\frac{\pi^2 j^2}{w^2}  \, \, ,
 \label{eq:qmu_multichannel}
\end{equation}
for an energy $E>\mu_j$ above the threshold, $j$ being a positive integer. In an energy window containing $N$ propagating channels, the wave function on link
$\vl$ can be written in a flux-normalized basis as
\begin{equation}
 \Psi_\vl(s,s_\perp)=
 \sum_{j=1}^{N}\sqrt{\frac{m}{\hbar q_j}}\,\Phi_j(s_\perp)
 \left(\alpha_{j,\vl}e^{iq_j s}
 +\beta_{j,\vl}e^{-iq_j s}\right) \, \, ,
 \label{eq:Psimulti}
\end{equation}
where $\alpha_{j,\vl}$ and $\beta_{j,\vl}$ are complex amplitudes. The analysis below concerns the scattering problem among those $N$ propagating-channels. Reflection symmetry about the axis of a straight link allows to separate these modes into 
$N_+$ even and $N_-$ odd channels, with $N=N_++N_-$.  We order the channel
basis according to this parity. 

We collect the propagation phases of the $N$ channels in the diagonal matrix
\begin{equation}
 K(E)=\operatorname{diag}
 \bigl(e^{iq_1(E)a_0},\ldots,e^{iq_N(E)a_0}\bigr) \, \, .
 \label{eq:multichannel_K_main}
\end{equation}
We denote $S_A(E)$ and $S_B(E)$ the unitary scattering matrices in $U(3N)$ that generalizes the vertex scattering matrices $U(3)$ of Sec.~\ref{sec:singlechannel} by including the transverse channel space. 

In the multichannel case, the generalization of Eq.~(\ref{detseculareqtrois}) reads
\begin{equation}
 \det[\mathbb I_{3N}-U(E,\vk)]=0 \, \, ,
 \label{eq:multichannel_secular_main}
\end{equation}
In the determinant, one round trip through the oriented unit cell is described by
\begin{equation}
 U(E,\vk)=
 (K(E)\otimes D_\vk^\dagger)S_B
 (K(E)\otimes D_\vk)S_A,
 \label{eq:multichannel_roundtrip_main}
\end{equation}
where $D_\vk$ is the diagonal matrix with diagonal entries $(1,e^{-i\vk\cdot \ve_1},e^{-i\vk \cdot \ve_2})$ as in Sec.~\ref{sec:singlechannel}, and the eventual energy dependence of $S_A$ and $S_B$ are implied.

\subsection{$D_3$-symmetric vertex matrix and the parity intertwiner
\label{sec:multichannel_vertex}}

We now analyse the structure of the scattering matrices $S_v$ using as a key ingredient the action of a reflection on the transverse
coordinate $s_\perp$ of each bond.  Let $C$ cyclically permute the three legs and let $R$ exchange
the two legs related by a chosen reflection. In the parity-ordered channel
basis, the physical representation of the two $D_3$ generators, $C_3$ and $M$, read 
\begin{equation}
  \rho(C_3) =\mathbb I_N\otimes C,  \hspace{2mm} {\rm and} \hspace{2mm}  \rho(M)=\Pi\otimes R,
 \label{eq:multichannel_D3_rep_main}
\end{equation}
where :
\begin{equation}
\Pi=\operatorname{diag}(\mathbb I_{N_+},-\mathbb I_{N_-}) \, \ ,
\end{equation}
acts differently on even vs odd states. Thus even channels carry the leg-space representation $(C,R)$, whereas odd
channels carry $(C,-R)$.  On the symmetric leg subspace
$\operatorname{Im}P_\parallel$, these sectors transform according to the
inequivalent one-dimensional irreducible representations $\mathbf 1_+$ and
$\mathbf 1_-$.  On $\operatorname{Im}P_\perp$, however, they give two
equivalent realizations of the same two-dimensional irreducible
representation, since $\mathbf 1_-\otimes\mathbf 2\simeq\mathbf 2$.
Consequently, $D_3$ symmetry allows opposite-parity channels to mix in the
$P_\perp$ sector.

To make this allowed mixing explicit, we first separate the incoming and
outgoing amplitudes according to transverse parity and write the corresponding block structure of the scattering matrix at $v$ :
\begin{equation}
 S_v=
 \begin{pmatrix}
  S_v^{++}&S_v^{+-}\\
  S_v^{-+}&S_v^{--}
 \end{pmatrix},
 \qquad v=A,B,
 \label{eq:multichannel_S_parity_blocks}
\end{equation}
where $S_v^{pp'}$ maps an incoming channel of parity $p'$ to an outgoing
channel of parity $p$.  A same-parity block acts between identical physical
leg representations.  Its action on $\operatorname{Im}P_\perp$ is therefore
proportional to the identity $P_\perp$ in that irreducible sector.  An
opposite-parity block instead maps leg amplitudes transforming under $(C,R)$
to amplitudes transforming under $(C,-R)$, or conversely.  If $J$ denotes the
leg-space part of this map, $D_3$ invariance requires
\begin{equation}
 CJ=JC,
 \qquad
 RJ=-JR.
 \label{eq:multichannel_intertwiner_conditions_main}
\end{equation}
The normalized solution 
\begin{equation}
 J=\frac{C-C^{-1}}{\sqrt3}.
 \label{eq:multichannel_J_main}
\end{equation}
acts on the three leg amplitudes, annihilates the symmetric leg vector,
and obeys $J^\dagger J=P_\perp$.  The channel-space matrices multiplying it
describe conversion between the corresponding even and odd transverse
modes. 

In the physical channel $\otimes$ leg basis, the most general parity-preserving blocks therefore read
\begin{equation}
 \label{eq:multichannel_S_main_same}
 \begin{aligned}
  S_v^{++}&=X_v^{++}\otimes P_\perp+Y_v^+\otimes P_\parallel \,\, ,\\
  S_v^{--}&=X_v^{--}\otimes P_\perp+Y_v^-\otimes P_\parallel \, \, ,
 \end{aligned}
\end{equation}
while the opposite-parity blocks are
\begin{equation}
 \label{eq:multichannel_S_main}
 \begin{aligned}
  S_v^{+-}&=X_v^{+-}\otimes J \, ,
  &S_v^{-+}&=X_v^{-+}\otimes J \, ,\\
 \end{aligned}
\end{equation}
These blocks describe conversion between even and odd channels. As we show below, this conversion generically gives the flat bands a finite width. Here $X_v^{pp'}$ maps incoming parity $p'$ to outgoing parity $p$.  The
matrices $Y_v^+$ and $Y_v^-$ act on the even and odd channel multiplicities
in the symmetric leg sector.  Unitarity requires
\begin{equation}
    \begin{pmatrix}X_v^{++}&-X_v^{+-}\\
X_v^{-+}&X_v^{--}\end{pmatrix} \in U(N)
\end{equation}
together with $Y_v^\pm\in U(N_\pm)$.  The derivation, the corresponding reciprocity conditions, and an explicit parametrization for one channel of each parity are given in Appendix~\ref{app:multichannel_parity}.

\subsection{Flat-band condition and parity-mixing obstruction
\label{sec:multichannel_flatness}}

We first consider the parity-conserving limit
\begin{equation}
 X_A^{+-}=X_A^{-+}=X_B^{+-}=X_B^{-+}=0 \, .
 \label{eq:multichannel_no_mixing_main}
\end{equation}
The two parity sectors then separate.  The leg vector
$\ket{\alpha_\vk}$ constructed in Eq.~\eqref{eq:alphak_singlechannel}
satisfies
$P_\perp\ket{\alpha_\vk}=\ket{\alpha_\vk}$ and
$P_\perp D_\vk\ket{\alpha_\vk}=D_\vk\ket{\alpha_\vk}$.  For parity
$p=+$ or $-$, the round-trip problem therefore reduces to the channel-space
condition
\begin{equation}
 \det\!\left[
 K_p(E)X_B^{pp}(E)K_p(E)X_A^{pp}(E)
 -\mathbb I_{N_p}\right]=0 \, ,
 \label{eq:multichannel_flat_condition_main}
\end{equation}
where $K_p$ is the restriction of $K$ to that parity sector.  This equation
contains no Bloch momentum.  Every one of its solutions consequently gives
an exact flat band.  Mixing among any number of propagating channels of the
same transverse parity does not invalidate the destructive-interference
construction.

The result changes when an opposite-parity block in
Eq.~\eqref{eq:multichannel_S_main} is present.  A parity-conversion event maps
$\ket{\alpha_\vk}$ to $J\ket{\alpha_\vk}$, which still belongs to
$\operatorname{Im}P_\perp$ at the same vertex.  Propagation to the neighboring
vertex does not preserve the second destructive-interference condition:
\begin{equation}
 P_\parallel D_\vk J\ket{\alpha_\vk}\ne0
 \qquad\text{at generic }\vk.
 \label{eq:multichannel_leakage_main}
\end{equation}
The converted amplitude therefore reaches the next vertex with a component
in the symmetric leg sector.  Since $P_\parallel$ and $P_\perp$ are acted on
by different channel matrices in Eq.~\eqref{eq:multichannel_S_main_same}, the
factorization leading to Eq.~\eqref{eq:multichannel_flat_condition_main} is
lost and the secular equation acquires Bloch-momentum dependence.  Local
$D_3$ symmetry thus permits the process that generically disperses the flat
band; it does not protect exact flatness when both transverse parities are
allowed to mix.

This obstruction also identifies the protected limits.  Exact flatness
survives when only one parity sector propagates, when an additional symmetry
forbids the blocks $X_v^{+-}$ and $X_v^{-+}$, or at exceptional parameter
values where the relevant parity-conversion amplitudes vanish or cancel.
For weak generic mixing, write the opposite-parity blocks as
$X_v^{+-},X_v^{-+}=O(\varepsilon)$, with no linear change in the
parity-preserving blocks.  A single insertion changes transverse
parity, so the first-order diagonal correction to an isolated
parity-conserving flat band vanishes.  Its leading dispersion is then
$O(\varepsilon^2)$.  If an even and an odd unperturbed band are degenerate,
degenerate perturbation theory can give a splitting at
$O(\varepsilon)$.  The complete perturbative expressions are derived in
Appendix~\ref{app:multichannel_parity}.

Fig.~\ref{fig:multichannel_parity_mixing} illustrates this mechanism in the
minimal multichannel problem with one propagating mode of each transverse
parity.  For $N_+=N_-=1$, the reciprocal vertex matrix can be parametrized by
an angle $\eta$ such that the opposite-parity amplitudes are proportional to
$\sin\eta$, while the same-parity amplitudes in
$\operatorname{Im}P_\perp$ are proportional to $\cos\eta$; the complete
parametrization is given in Appendix~\ref{app:multichannel_parity}.  We use
identical $D_3$-symmetric matrices at the two vertices and hold all their
remaining parameters fixed.  At $\eta=0$, the even and odd sectors decouple
and the two highlighted solutions are exact flat bands.  Activating $\eta$
mixes their transverse character and gives both descendants a visible
dispersion.

\begin{figure}[!htbp]
 \centering
 \includegraphics[width=\columnwidth]{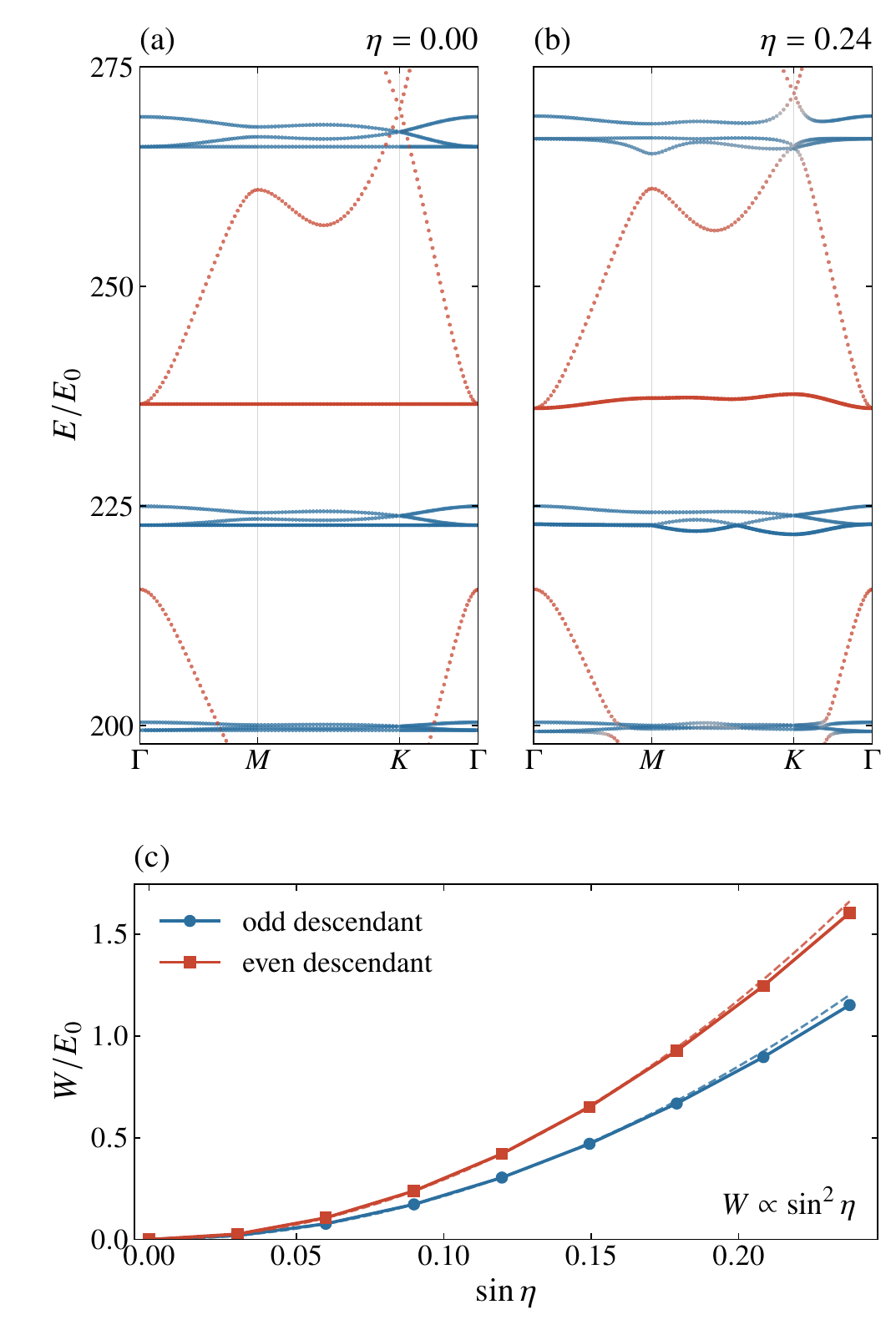}
 \caption{Parity mixing in a honeycomb quantum graph with one propagating
 even channel and one propagating odd channel, $N_+=N_-=1$.  The displayed
 energy window lies above the second transverse threshold and below the
 third, so exactly these two transverse modes propagate.  The spectra in
 panels (a) and (b) are shown along $\Gamma$--$M$--$K$--$\Gamma$ and contain
 several longitudinal band families.  The color of each point gives its
 transverse-channel character: blue and red denote odd and even character,
 respectively, while intermediate colors indicate mixing. The angle $\eta$
 controls the opposite-parity scattering amplitudes, whose magnitude is
 $\sin\eta$ in the $\operatorname{Im}P_\perp$ sector. The thick curves
 track one flat-band descendant from each parity sector.  (a) In the
 parity-conserving limit $\eta=0$, both tracked bands are
 exactly flat.  (b) For $\eta=0.24$, opposite-parity scattering through the
 $J$ blocks gives both descendants a finite width.  All parameters other than
 $\eta$ are unchanged. (c) Bandwidths $W$ of the two tracked descendants
 as functions of the parity-conversion amplitude $\sin\eta$. Blue circles
 and red squares denote the odd- and even-sector descendants, respectively.
 Each width is the energy range over the Brillouin zone. The descendant
 bands are tracked continuously from $\eta=0$, as described in
 Appendix~\ref{app:multichannel_parity}. The dashed curves
 are quadratic guides obtained from the four smallest nonzero mixing
 amplitudes. We use identical, energy-independent reciprocal
 vertices with $w/a_0=0.45$ and
 $(\gamma,\alpha,\beta,\theta_+,\theta_-)
 =(1.50,0.50,\pi/2,-0.35,0.80)$ in the parametrization of
 Appendix~\ref{app:multichannel_parity};
 $E_0=\hbar^2/(2ma_0^2)$.}
 \label{fig:multichannel_parity_mixing}
\end{figure}

Fig.~\ref{fig:multichannel_parity_mixing}(c) shows that both descendant bandwidths are consistent with the predicted quadratic scaling at weak mixing.

In the antidot regime of Sec.~\ref{sec:2DEGantidots}, only the lowest
transverse mode propagates over the relevant energy window. A junction can
excite a higher, evanescent transverse mode, which decays along the trench.
When this mode has odd parity, its conversion with the even propagating mode
is a counterpart of the even--odd conversion studied here, where both modes
propagate. Tunneling through the finite antidot barriers provides an additional
transfer path omitted from this graph.

\section{Conclusion \label{sec:conclusion}}  

We have investigated the spectral properties of two classes of systems: (i) a two-dimensional electron gas (2DEG) under a periodic electrostatic antidot potential and (ii) quantum-graph models in both single-channel and multichannel regimes.

In the high-barrier, narrow-trench regime of the 2DEG antidot model (Sec.~\ref{sec:2DEGantidots}), the low-energy spectrum contains two distinct families of bands. The two {\it graphene-like} bands ($n=1,2$) arise from hybridization of local junction states and are well described by a two-orbital honeycomb model, with junction-centered orbitals. The higher {\it kagome-like} triplets combine a nearly flat band with two dispersive partners, all with wave-function weight concentrated along the trenches. Their dispersion reflects coupling at the junctions and through the finite antidot barriers. The nearly flat members $n=3,6,9,\ldots$ evolve toward the exact flat bands of the quantum-graph model as the single-channel limit is approached.

In its singlechannel version (Sec. \ref{sec:singlechannel}), the quantum graph model harbors exact flat bands extending other the whole Brillouin zone. Coexisting with 2 generically dispersive bands, each of these FB is supported by CLS and can be seen as the limit, or an idealized modelization, of the nearly FB of model i) for strongly confined, single-mode trenches. In this extreme limit, which is nevertheless realized in some real systems, we have provided a general argument to explain the origin and robustness of the FBs, both in reciprocal and real spaces.

The multichannel version (Sec. \ref{sec:multichannel}) realizes somewhat an intermediate limit between the model i) and the singlechannel model ii). In the ideal multichannel graph, same-parity mixing preserves the flat-band construction, whereas even–odd conversion generically produces dispersion. The latter provides a propagating-channel counterpart of the evanescent transverse-mode conversion discussed for the antidot model. Direct tunneling through the antidot barriers introduces additional couplings that are omitted from the graph considered here.

These results are directly relevant to periodically gated or etched high-mobility 2DEGs,
molecule-patterned metallic surfaces, and ballistic waveguide networks. They
also provide a controlled starting point for studying interaction and disorder
effects once the exact graph degeneracy is weakly broadened in a realistic
device.

\medskip

{\it Acknowledgements.} We thank Jean-Noël Fuchs, Dario Bercioux and Vincent Bruneau for insightful discussions about the antidot model. JC thanks Joel Moore and UC Berkeley physics department for hospitality, in the framework of the Fulbright program 2025. JC was supported by the RIE program "Flatbands" of Bordeaux University. CL was supported by the U.S. Department of Energy, Office of Science, National Quantum Information Science Research Centers, Quantum Science Center. 

\appendix

\onecolumngrid

\section{Numerical construction of the antidot spectra and observables
\label{app:antidot_numerics}}

The continuum eigenvalues and eigenfunctions provide the spectra, density
plots, and parameter maps in Figs.~\ref{fig:antidot_model_states},
\ref{fig:antidot_graph_crossover}, \ref{fig:antidot_junction_splitting},
and \ref{fig:antidot_bandwidth_map}. A separate continuum scattering problem
for one Y junction supplies the single-channel graph overlay in
Fig.~\ref{fig:antidot_graph_crossover}(c). In that calculation we first
solve for the transverse trench modes, use them to impose incoming and
outgoing waves at the junction boundary, and extract the outgoing
amplitudes from the resulting junction solutions. These amplitudes fix
the scattering matrix used in the graph calculation.

Throughout the appendices we set $\hbar^2/(2m)=1$; for the continuum calculations in this appendix and Appendix~\ref{app:triangle_junction}, we also set $a_0=1$.

\subsection{Periodic spectra and probability densities}

For the geometry of Sec.~\ref{sec:2DEGantidots} in these units, the antidot side length is
$h=1-w/\sqrt3$, the primitive-cell area is $A=3\sqrt3/2$, and the
reciprocal vectors are $\vb_1=(2\pi/\sqrt3,2\pi/3)$ and
$\vb_2=(-2\pi/\sqrt3,2\pi/3)$.
For band index $n$, the normalized Bloch state is reconstructed as
\begin{equation}
 \psi_{n\vk}(\vr)=\frac{e^{i\vk\cdot\vr}}{\sqrt A}
 \sum_{\vG\in\mathcal C_{M_h}}c_{n\vG}(\vk)e^{i\vG\cdot\vr},
 \qquad \sum_{\vG}|c_{n\vG}|^2=1.
 \label{eq:A_bloch_reconstruction}
\end{equation}
Here $\vG=m_1\vb_1+m_2\vb_2$, with integer reciprocal indices
$m_1,m_2$. The point-group-invariant basis is
\begin{equation}
 \mathcal C_{M_h}=\{\vG:
 \max(|m_1|,|m_2|,|m_1-m_2|)\le M_h\}.
 \label{eq:hexagonal_reciprocal_cutoff}
\end{equation}
It contains $1+3M_h(M_h+1)$ plane waves and is held fixed throughout each
momentum calculation.

The potential matrix in Eq.~\eqref{h_mat} is obtained from the Fourier
transform of a single antidot hexagon $\mathcal H_h$:
\begin{equation}
 V_{\vG\vG'}=\frac{V_0}{A}F_h(\vG-\vG'),\qquad
 F_h(\vq)=\int_{\mathcal H_h}e^{-i\vq\cdot\vr}\,d^2r.
 \label{eq:antidot_potential_matrix_app}
\end{equation}
We evaluate the integral along the six edges. Let
$\vr_j=h(\cos\alpha_j,\sin\alpha_j)$, with
$\alpha_j=\pi/2+(j-1)\pi/3$, be the counterclockwise vertices, and set
$\vr_7=\vr_1$ and $\vd_j=\vr_{j+1}-\vr_j$. Green's theorem gives,
for $\vq\ne0$,
\begin{equation}
 F_h(\vq)=\frac{i}{|\vq|^2}\sum_{j=1}^{6}
 (q_xd_{j,y}-q_yd_{j,x})e^{-i\vq\cdot\vr_j}
 g(\vq\cdot\vd_j),
 \label{eq:hexagon_fourier_boundary}
\end{equation}
where $g(x)=(1-e^{-ix})/(ix)$, $g(0)=1$, and
$F_h(0)=3\sqrt3h^2/2$. The lowest eigenpairs of the resulting Hermitian
Bloch matrix are obtained by iterative diagonalization; the large parameter
scan uses matrix-free application of the Fourier convolution.

For Figs.~\ref{fig:antidot_model_states}(b) and
\ref{fig:antidot_graph_crossover}, we solve along
$\Gamma$--$M$--$K$--$\Gamma$, taking
$M=\vb_1/2$ and $K=(2\vb_1+\vb_2)/3$, with 41 equal momentum
steps per segment. The band labels follow increasing energy at each
momentum. Every branch entering the displayed energy interval is retained.
Fig.~\ref{fig:antidot_model_states} uses $M_h=92$; the three panels
of Fig.~\ref{fig:antidot_graph_crossover} use $M_h=56,65,92$,
respectively.

Equation~\eqref{eq:A_bloch_reconstruction} also gives the two density
diagnostics in Fig.~\ref{fig:antidot_model_states}(c,d):
\begin{equation}
 \rho_{12,\Gamma}(\vr)=\sum_{n=1}^{2}|\psi_{n\Gamma}(\vr)|^2,\qquad 
 \rho_{3,M\star}(\vr)=\frac{1}{3}\sum_{j=1}^{3}|\psi_{3M_j}(\vr)|^2.
 \label{eq:A_density_diagnostics}
\end{equation}
The first density probes the {\it graphene-like} bands and is independent of basis rotations within their two-state subspace. The second probes the lowest nearly flat {\it kagome-like} band.
For the second, we choose the three symmetry-related $M_j$ because band 3
is separated there from neighboring bands, avoiding its degeneracy with
band 4 at $\Gamma$. Rotating the density at one $M$ point generates the
other two terms. Their equal-weight average restores threefold symmetry;
it is an incoherent density average, not a single Bloch eigenstate.
Each surface is divided by its own maximum. The hexagonal display window,
of circumradius $3a_0/2$ and centered on an antidot, only clips the
reconstructed periodic density; it imposes no boundary condition on the
eigenproblem.

\subsection{Straight-trench modes and the local junction domain}

The energy origin in Fig.~\ref{fig:antidot_graph_crossover} is the lowest
transverse eigenenergy $\mu_1(V_0,w)$ of an infinite straight trench.
Its transverse potential is
\begin{equation}
 V_\perp(y)=
 \begin{cases}
 0,&|y|<w/2,\\
 V_0,&|y|>w/2.
 \end{cases}
 \label{eq:A_transverse_potential}
\end{equation}
We solve $[-\partial_y^2+V_\perp(y)]\chi=\mu\chi$ on the whole
real line, with decay as $|y|\to\infty$. Matching the even ground state
and its derivative at $y=\pm w/2$ gives
\begin{equation}
 p_1=\frac{w\sqrt{V_0}}{2}\cos p_1,\qquad
 \mu_1=\left(\frac{2p_1}{w}\right)^2,
 \quad 0<p_1<\frac\pi2.
 \label{eq:A_trench_threshold}
\end{equation}
Here $p_1=w\sqrt{\mu_1}/2$. The same independently calculated $\mu_1$
is subtracted from the continuum and graph energies in each panel.

\begin{figure}[htbp]
 \centering
 \includegraphics[width=0.97\linewidth]{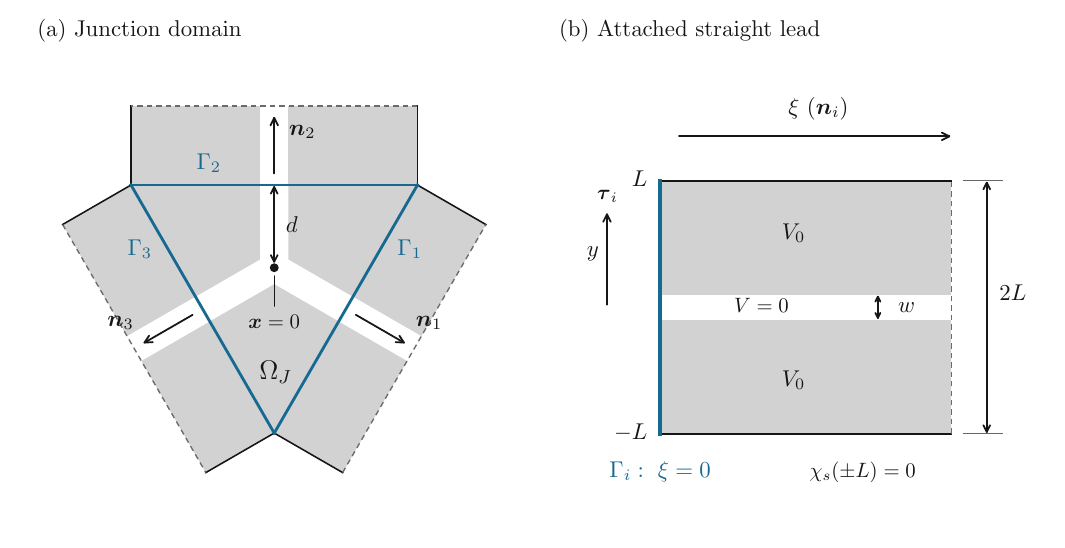}
 \caption{Junction domain and lead coordinates for the local
 scattering calculation.
 (a) The three blue interfaces $\Gamma_i$ bound the full two-dimensional
 region $\Omega_J$, including both barrier material (gray, $V=V_0$) and
 trenches (white, $V=0$). The junction center $\boldsymbol{x}=0$ lies a
 perpendicular distance $d$ from each interface. The geometric ratios are
 those used for Fig.~\ref{fig:antidot_graph_crossover}(c).
 (b) Local coordinates in one lead: $\xi$ increases outward along
 $\boldsymbol n_i$, and $y$ runs across the lead along $\boldsymbol\tau_i$.
 The trench width $w$ differs from the full numerical lead width
 $2L=2\sqrt3d$. Transverse profiles $\chi_s(y)$ are solved on $[-L,L]$
 with Dirichlet endpoints; propagation or decay occurs along $\xi$.
 Junction and leads meet at $\xi=0$ without overlapping interiors.
 Dashed ends indicate display truncations of the semi-infinite leads,
 not additional boundaries.}
 \label{fig:antidot_junction_domain}
\end{figure}

To determine the junction scattering matrix, we retain the two-dimensional
antidot potential in a finite region around one junction and attach three
straight leads. Fig.~\ref{fig:antidot_junction_domain} shows the
junction domain and the coordinates in the attached leads.
Choose the junction at $\vr_J=(0,1)$, relative to the
antidot center at the origin, and write $\boldsymbol{x}=\vr-\vr_J$.
The three outward trench axes are
\begin{equation}
 \boldsymbol n_1=(\sqrt3/2,-1/2),\qquad \boldsymbol n_2=(0,1), \qquad
 \boldsymbol n_3=(-\sqrt3/2,-1/2).
 \label{eq:A_junction_axes}
\end{equation}
For a chosen matching distance $d$, the junction region is the
$D_3$-symmetric equilateral triangle
\begin{equation}
 \Omega_J=\{\boldsymbol{x}:\boldsymbol n_i\cdot\boldsymbol{x}<d,
 \ i=1,2,3\},\qquad
 V_J(\boldsymbol{x})=V(\vr_J+\boldsymbol{x}).
 \label{eq:A_junction_domain}
\end{equation}
Its vertices are $(0,-2d)$ and $(\pm\sqrt3d,d)$. Thus $V_J$ is
the restriction of the original antidot potential, including both the
barrier regions and the trench portions inside the triangle.

Let $\boldsymbol\tau_i=(-n_{i,y},n_{i,x})$ be the transverse unit vector.
The side normal to $\boldsymbol n_i$ is the one-dimensional interface
\begin{equation}
 \Gamma_i=\{d\boldsymbol n_i+y\boldsymbol\tau_i:-L<y<L\},
 \qquad L=\sqrt3d.
 \label{eq:A_junction_interfaces}
\end{equation}
The attached lead consists of points
$(d+\xi)\boldsymbol n_i+y\boldsymbol\tau_i$, with $\xi>0$ and
$|y|<L$. Its potential is $V_\perp(y)$, independent of the outward
longitudinal coordinate $\xi$, and its sidewalls at $y=\pm L$ carry
Dirichlet conditions. The junction and leads have disjoint interiors
and join at $\xi=0$ along $\Gamma_i$. The transverse modes are functions
on these one-dimensional interfaces and are continued analytically into
the leads; the two-dimensional junction field is solved only in
$\Omega_J$. Straight leads replace
the surrounding periodic network outside these interfaces.
At the overlay parameters, the cuts lie in straight trench portions,
and the restriction of $V_J$ to each cut is $V_\perp(y)$.

The boundary basis is obtained from the trench problem on this finite
transverse interval:
\begin{equation}
 [-\partial_y^2+V_\perp(y)]\chi=\lambda\chi,
 \qquad \chi(-L)=\chi(L)=0.
 \label{eq:A_transverse_dirichlet_problem}
\end{equation}
Here $\lambda$ denotes a transverse eigenenergy. We solve this problem
with continuous piecewise-linear finite elements on the same line nodes
as the junction boundary. For each interior node $y_a$, the basis
function $\zeta_a(y)$ equals one at $y_a$ and zero at all other mesh nodes,
including the endpoints. The matrices are
\begin{equation}
 M^\perp_{ab}=\int_{-L}^{L}\,\zeta_a\zeta_b \,dy,\qquad
 K^\perp_{ab}=\int_{-L}^{L}\, \zeta'_a\zeta'_b\,dy,\qquad
 V^\perp_{ab}=\int_{-L}^{L}\,\zeta_a V_\perp\,\zeta_b\,dy.
 \label{eq:A_transverse_matrices}
\end{equation}
The numerical eigenpairs used below are defined by
\begin{equation}
 \begin{aligned}
 (K^\perp+V^\perp)c_s&=\widetilde\mu_s M^\perp c_s,\\
 \chi_s(y)&=\sum_a(c_s)_a\,\zeta_a(y),\qquad
 c_s^\dagger M^\perp c_t=\delta_{st}.
 \end{aligned}
 \label{eq:A_discrete_transverse_modes}
\end{equation}
We label the modes in increasing eigenenergy. Their normalization gives
$\int_{-L}^{L}\chi_s^*\chi_t\,dy=\delta_{st}$. Thus
$\widetilde\mu_1$ is the lowest eigenvalue of the \emph{discretized,
finite-interval} trench problem. It differs from the whole-line
threshold $\mu_1$ through both transverse truncation and finite-element
error. For a bound transverse state, these errors vanish as the interval
is enlarged and the mesh is refined. The ground mode $\chi_1$ is even;
its sign is chosen positive and identically in all three leads.

For Fig.~\ref{fig:antidot_graph_crossover}(c), we use $V_0/E_0=400$,
$w/(\sqrt3a_0)=0.05$, and $d/a_0=0.25125$. The equilateral mesh
has edge length $\delta=w/20$: its nodes are
$\delta(m+n/2,\sqrt3n/2)$ inside the closed triangle, with neighboring
nodes joined into elementary triangles. The three corner values vanish,
consistently with the lead sidewalls. This leaves 20\,500 junction
unknowns and 200 transverse modes on each interface. The potential
$V_J$ on each triangle is sampled at its centroid, multiplied by the
exact element mass matrix, and averaged over the three rotations to
preserve symmetry. The integrals in $V^\perp$ are evaluated with the
steps at $y=\pm w/2$ resolved exactly. These choices give
$\widetilde\mu_1/E_0\simeq243.603$ and
$\widetilde\mu_2/E_0\simeq448.726$, whereas
$\mu_1/E_0\simeq243.520$.

\subsection{Boundary matching, scattering amplitudes, and graph overlay}

At a specified scattering energy $E$ between
$\widetilde\mu_1$ and $\widetilde\mu_2$, only $\chi_1$ propagates.
Its longitudinal wave number and the decay constants of the remaining
modes are determined by the already calculated transverse energies:
\begin{equation}
 q(E)=\sqrt{E-\widetilde\mu_1},\qquad
 \alpha_s(E)=\sqrt{\widetilde\mu_s-E}\quad(s>1).
 \label{eq:A_lead_wave_numbers}
\end{equation}
Within this transverse discretization, a boundary value
$f(y)=\sum_s f_s\chi_s(y)$ has the outgoing lead continuation
$f_1\chi_1(y)e^{iq\xi}+\sum_{s>1}f_s\chi_s(y)e^{-\alpha_s\xi}$.
Differentiating this known continuation at $\xi=0$ defines
\begin{equation}
 \mathcal T_{\mathrm{out}}(E)f
 =iq f_1\chi_1-\sum_{s>1}\alpha_s f_s\chi_s,\qquad
 f_s=\int_{-L}^{L}\chi_s^*(y)f(y)\,dy.
 \label{eq:A_outgoing_operator}
\end{equation}
This outgoing Dirichlet-to-Neumann operator maps a boundary value to the
normal derivative of its outgoing or decaying lead continuation.
For given $(V_0,w,d,E)$, the transverse eigenpairs and longitudinal
wave numbers in Eqs.~\eqref{eq:A_discrete_transverse_modes} and
\eqref{eq:A_lead_wave_numbers} determine $\mathcal T_{\mathrm{out}}$.
All 200 transverse modes are retained in this boundary operator.
Modes with $\widetilde\mu_s>E$ are evanescent along the lead,
whether or not they correspond to transversely bound states on the
untruncated line.

To find the response to a unit incoming wave in lead $j$, write the field
in lead $i$ as
\begin{equation}
 \psi_i^{(j)}(\xi,y)=\delta_{ij}\chi_1(y)e^{-iq\xi}
 +(S_v^{\mathrm{port}})_{ij}\chi_1(y)e^{iq\xi}+\sum_{s>1}b_{is}^{(j)}\chi_s(y)e^{-\alpha_s\xi}.
 \label{eq:A_incident_lead_expansion}
\end{equation}
Here $\delta_{ij}$ fixes the prescribed incident amplitude;
$(S_v^{\mathrm{port}})_{ij}$ and $b_{is}^{(j)}$ are the unknown
outgoing propagating and evanescent amplitudes. The superscript
``port'' specifies that amplitudes are referred to the matching
cross-sections $\xi=0$. The corresponding unknown junction field is
$\Psi_j(\boldsymbol{x})$.

The lead expansion and the junction solution describe the same
wavefunction on opposite sides of $\Gamma_i$. Continuity of the
wavefunction and its normal derivative therefore requires
\begin{equation}
 \begin{aligned}
 \Psi_j(d\boldsymbol n_i+y\boldsymbol\tau_i)
 &=(\delta_{ij}+(S_v^{\mathrm{port}})_{ij})\chi_1
   +\sum_{s>1}b_{is}^{(j)}\chi_s,\\
 \partial_{n_i}\Psi_j\big|_{\Gamma_i}
 &=iq((S_v^{\mathrm{port}})_{ij}-\delta_{ij})\chi_1
   -\sum_{s>1}\alpha_s b_{is}^{(j)}\chi_s.
 \end{aligned}
 \label{eq:A_interface_matching}
\end{equation}
The normal points out of $\Omega_J$ into the lead, so
$\partial_{n_i}=\partial_\xi$ at the interface. Subtracting
$\mathcal T_{\mathrm{out}}\Psi_j|_{\Gamma_i}$ from the second line
eliminates all unknown outgoing amplitudes and leaves the known source
$-2iq\delta_{ij}\chi_1$. Thus the finite-domain problem to solve is
\begin{equation}
 \begin{aligned}
 (-\nabla^2+V_J-E)\Psi_j&=0&&\text{in }\Omega_J,\\
 (\partial_{n_i}-\mathcal T_{\mathrm{out}})\Psi_j
 &=-2iq\delta_{ij}\chi_1&&\text{on }\Gamma_i.
 \end{aligned}
 \label{eq:A_junction_scattering_problem}
\end{equation}
The sign and factor of two follow from the derivative $-iq$ of the
incident wave and $+iq$ of the outgoing wave. The boundary values of
$\Psi_j$ are solved for, not prescribed independently alongside their
derivatives.

The finite-element form of Eq.~\eqref{eq:A_junction_scattering_problem}
follows by multiplying by a two-dimensional finite-element test function
$\zeta^*(\boldsymbol{x})$ and integrating by parts:
\begin{equation}
 \int_{\Omega_J}\!\bigl[\nabla\zeta^*\cdot\nabla\Psi_j   +(V_J-E)\zeta^*\Psi_j\bigr]d^2x
 \quad-\sum_i\int_{\Gamma_i}\!\zeta^*\mathcal T_{\mathrm{out}}    (\Psi_j|_{\Gamma_i})\,dy
   =-2iq\int_{\Gamma_j}\!\zeta^*\chi_1\,dy.
 \label{eq:A_junction_weak_form}
\end{equation}
We use the piecewise-linear basis on the triangular mesh specified
above. If $\Phi$ collects the transverse eigenvectors $c_s$ and
$D_{\mathrm{out}}=\operatorname{diag}(iq,-\alpha_2,\ldots)$,
each boundary contribution to the matrix on the left is
$-M^\perp\Phi D_{\mathrm{out}}\Phi^\dagger M^\perp$; the incident boundary
load is $-2iqM^\perp c_1$ on lead $j$. The same sparse matrix serves
the three right-hand sides $j=1,2,3$.

After each solve, projecting the first line of
Eq.~\eqref{eq:A_interface_matching} onto $\chi_1$ removes the higher
transverse modes by orthogonality and gives
\begin{equation}
 (S_v^{\mathrm{port}})_{ij}
 =\int_{-L}^{L}\chi_1^*(y)
   \Psi_j(d\boldsymbol n_i+y\boldsymbol\tau_i)\,dy-\delta_{ij}.
 \label{eq:A_scattering_projection}
\end{equation}
The integral is the \emph{total} fundamental-mode amplitude at the
interface; subtracting the known incoming amplitude leaves the outgoing
one. Column $j$ is consequently the response to incidence in lead $j$.
By linearity, an arbitrary incoming vector $a^{\mathrm{in}}$ produces
$a^{\mathrm{out}}=S_v^{\mathrm{port}}a^{\mathrm{in}}$, which establishes
that this projection constructs the scattering matrix. The transverse
mode $\chi_1$ supplies the channel basis, while $\Psi_j$ contains the
response of the two-dimensional junction. All three propagating channels
have the same $q$, so this amplitude matrix is also flux normalized.
For a real potential, flux conservation makes it unitary; the evanescent
components carry no flux at infinity.

The even fundamental modes and $D_3$ symmetry give common diagonal and
off-diagonal entries $r_v^{\mathrm{port}}$ and $t_v^{\mathrm{port}}$.
The two scattering phases of Sec.~\ref{sec:singlechannel} follow from
\begin{equation}
 \begin{aligned}
 e^{i\phi_v^{\mathrm{port}}}&=r_v^{\mathrm{port}}-t_v^{\mathrm{port}},\\
 e^{i\theta_v^{\mathrm{port}}}&=r_v^{\mathrm{port}}+2t_v^{\mathrm{port}}.
 \end{aligned}
 \label{eq:A_scattering_phases}
\end{equation}
For the specified matching geometry, we determine the phases as functions
of $(E,w,V_0)$ from the isolated-junction scattering problem, independently
of the periodic continuum spectrum. We calculate them at 13 equally
spaced energies with $4\le(E-\mu_1)/E_0\le38$, interpolate the two
complex eigenvalues with cubic splines, and restore unit modulus after
interpolation. With identical vertices $S_A=S_B$, the graph spectrum is
then obtained from the secular equation of Sec.~\ref{sec:singlechannel}.
Numerically we track its round-trip eigenphases on 801 equally spaced
energies over this interval and interpolate their crossings of integer
multiples of $2\pi$.

The reference planes fix the propagation length. With
$S_v^{\mathrm{port}}$ it is $\ell=a_0-2d$; with the center-referenced
matrix and length $a_0$ the equivalent prescription is
\begin{equation}
 S_v^{\mathrm{center}}(E)=e^{-2iq(E)d}S_v^{\mathrm{port}}(E).
 \label{eq:A_scattering_reference_planes}
\end{equation}
We use the same $\widetilde\mu_1$ for junction scattering and link
propagation. At the displayed flat-band root
$E_\star/E_0\simeq254.930$, the interpolated port phases are
$(\phi_v^{\mathrm{port}},\theta_v^{\mathrm{port}})
\simeq(-1.674,-2.980)$ radians. Only the lowest propagating mode is
carried between graph vertices; the evanescent modes enter the local
scattering calculation. This reduction supplies the dashed spectrum,
while the finite continuum bandwidth is calculated directly below.

\subsection{Parameter maps, junction hopping, and numerical resolution}

For each $(V_0,w)$ in Figs.~\ref{fig:antidot_junction_splitting} and
\ref{fig:antidot_bandwidth_map}, we compute
the four lowest continuum eigenvalues on a centered hexagonal
Brillouin-zone mesh. Fig.~\ref{fig:antidot_junction_splitting} uses the splitting
$\Delta_{\mathrm{BS}}=E_2(\Gamma)-E_1(\Gamma)$, and
Fig.~\ref{fig:antidot_bandwidth_map} uses
Eq.~\eqref{eq:Wn_fullBZ} for $n=3$. Both quantities are evaluated from
the calculated eigenvalues before interpolation of the parameter maps.
The two maps share a grid of 91 widths and 37 barrier heights and the
fixed reciprocal shell $M_h=96$. The barrier heights run from
$100E_0$ to $1000E_0$ in steps of $25E_0$. The width ratio
$w/(\sqrt3a_0)$ is sampled with steps $0.0025$, $0.005$, and $0.01$
on the intervals $[0.025,0.10]$, $[0.10,0.30]$, and $[0.30,0.50]$,
respectively. The momentum mesh consists of
$\vk=(j_1\vb_1+j_2\vb_2)/N_k$, with integers $j_1,j_2$ obeying
$\max(|2j_1-j_2|,|2j_2-j_1|,|j_1+j_2|)\le N_k$.
For $N_k=18$ it contains 343 points, including $\Gamma$, $M$, and $K$;
point-group symmetry reduces these to 37 independent calculations.

In order to test the two-orbital description of the {\it graphene-like} bands in Eq.~\eqref{eq:junction_two_orbital},
we fit the splitting over the sampled Brillouin zone to
$2|t_J||f(\vk)|$, where
$f(\vk)=1+e^{i\vk\cdot\ve_1}+e^{i\vk\cdot\ve_2}$.
Minimizing the equal-weight sum of squared residuals gives
\begin{equation}
 |t_J|_{\mathrm{fit}}=
 \frac{\sum_\vk |f(\vk)|[E_2(\vk)-E_1(\vk)]}{2\sum_\vk |f(\vk)|^2}.
 \label{eq:A_hopping_extraction}
\end{equation}
At 18 tested parameter points with $V_0/E_0=200,400,800$ and
$w/(\sqrt3a_0)=0.025,0.05,0.10,0.20,0.30,0.50$, the root-mean-square
residual is $0.2$--$4.2\%$ of $\Delta_{\mathrm{BS}}$.
In the narrow, high-barrier regime the fitted magnitude agrees with
$\Delta_{\mathrm{BS}}/6$ within $0.5\%$; deviations increase for wider
trenches. These fits test the interpretation of the directly calculated $\Gamma$-point splitting in Fig.~\ref{fig:antidot_junction_splitting} as $6t_J$. The triangular-junction model supplies the separate approximate prediction compared with the continuum results in Fig.~\ref{fig:triangle-junction}. The relation
$W_1=W_2=\Delta_{\mathrm{BS}}/2$ retains the nearest-neighbor
approximation stated in the main text.

The numerical checks distinguish absolute energy accuracy from the
relative accuracy of a small bandwidth. Higher-shell checks at the
vertices of the spectral path change the displayed energies by less
than $9\times10^{-4}E_0$; increasing $M_h$ from 92 to 104 changes the
normalized density surfaces by at most $2.1\times10^{-4}$.
For the parameter maps, representative cutoff tests support
$\Delta_{\mathrm{BS}}$ and the region with
$W_3\gtrsim5\times10^{-3}E_0$ at the resolution of the figure.
Increasing $N_k$ from 18 to 30 at six check points changes $W_3$ by
less than $3\times10^{-8}E_0$; basis truncation is the more restrictive
control on very small widths.

In the narrow-band valley, the two band edges converge while their much
smaller difference can remain sensitive to the reciprocal cutoff.
Consequently, values below $5\times10^{-3}E_0$ locate the valley but
are not assigned quantitative relative accuracy. The lower color limit
$10^{-4}E_0$ in Fig.~\ref{fig:antidot_bandwidth_map} is a separate
display choice: all smaller values share the lowest color. At the
smallest trench widths, cutoff sensitivity prevents a reliable
determination of whether $W_3$ increases again.

\section{Analytical triangular junction model}
\label{app:triangle_junction}

We model the potential-free region around a $Y$ junction by an equilateral triangle connected to three equivalent straight trenches. The lowest junction-centered state belongs to the fully symmetric representation of the local $D_3$ symmetry, so we restrict the analysis to this sector.

A convenient analytical form for the symmetric mode of an equilateral triangular resonator is
\begin{equation}
    \chi_{p,\varphi}(\mathbf r) =
    \sum_{j=1}^{3}
    \cos\!\left[p\,d_j(\mathbf r)+\varphi\right],
\end{equation}
where $d_j(\mathbf r)$ is the perpendicular distance from the point $\mathbf r$ inside the triangle to side $j$ of the triangle. The parameter $p$ plays the role of the wave number inside the triangular core, while $\varphi$ controls the phase of the standing-wave pattern relative to the boundaries.

Because each distance function satisfies $|\nabla d_j|=1$, every term in the sum is an eigenfunction of the Laplacian with eigenvalue $p^2$. Therefore

\begin{equation}
    -\nabla^2 \chi_{p, \varphi} = 
    p^2 \chi_{p, \varphi},
\end{equation}
and the energy inside the junction core is simply $E=p^2$.

We approximate the response of the three outgoing trenches by the same scalar boundary condition on each triangle side :
\begin{equation}
    \partial_n \Psi + \beta(E) \Psi = 0,
\end{equation}
where $\partial_n$ is the outward normal derivative and $\beta(E)$ is the logarithmic derivative imposed by the longitudinal trench wavefunction. This reduced Robin model applies the logarithmic derivative of the lowest trench mode to the entire side, without requiring the junction boundary trace to have that mode's transverse profile. Higher-mode matching and direct coupling through the antidots are omitted. Substituting the analytical triangular mode into this condition and separating the constant and position-dependent terms along one side gives two algebraic relations for $p$ and $\varphi$.
Eliminating $\varphi$ yields the scalar secular equation

\begin{equation}
    F_\triangle(p,\beta)=0,
\end{equation}
with
\begin{equation}
    F_\triangle(p,\beta) =
    \left(p^2-2\beta^2\right)
    \sin\!\left(\frac{pH}{2}\right) -
    3\beta p
    \cos\!\left(\frac{pH}{2}\right),
    \label{eq:triangle_secular}
\end{equation}
where $H=\sqrt3w/2$ is the height of the equilateral triangle of side $w$.

We now couple the triangular junction to the lowest transverse mode of the neighboring trench. Let $\mu_1$ be the lowest transverse threshold of the straight trench. For a junction state below this threshold, $E < \mu_1$, the longitudinal trench mode is evanescent with decay constant $\gamma(E) = \sqrt{\mu_1 - E}$.

To describe the {\it graphene-like} bands, neighboring triangular junctions are connected by a finite trench segment. At the $\Gamma$ point the junction amplitudes can be chosen either symmetric or antisymmetric. For $E<\mu_1$, solving the one-mode link problem gives the boundary coefficients

\begin{equation}
    \beta_+(E) =
    \gamma(E)
    \tanh\!\left[\frac{\gamma(E)h}{2}\right],
    \qquad
    \beta_-(E) =
    \gamma(E)
    \coth\!\left[\frac{\gamma(E)h}{2}\right].
\end{equation}

If a root crosses above $\mu_1$, the longitudinal trench amplitude becomes oscillatory. Within the same one-mode approximation, continuation with $\gamma=iq_{\mathrm t}$, where $q_{\mathrm t}=\sqrt{E-\mu_1}$, gives
\begin{equation*}
\begin{aligned}
    \beta_+(E)&=-q_{\mathrm t}\tan\!\left(\frac{q_{\mathrm t}h}{2}\right),\\
    \beta_-(E)&=q_{\mathrm t}\cot\!\left(\frac{q_{\mathrm t}h}{2}\right),
    \qquad E>\mu_1.
\end{aligned}
\end{equation*}

At $E=\mu_1$, the continuous limits are $\beta_+=0$ and $\beta_-=2/h$. These expressions cover the parameter range of Fig.~\ref{fig:triangle-junction}, including cases where the antibonding state lies above threshold.

The bonding and antibonding energies $E_b$ and $E_{ab}$ are the lowest positive roots of Eq.~\eqref{eq:triangle_secular} with $\beta_+$ and $\beta_-$, respectively, continued through the threshold when necessary. Finally, identifying these two roots with the two lowest $\Gamma$-point states of the effective honeycomb two-orbital model gives

\begin{equation}
    \epsilon_J^\triangle = \frac{E_b + E_{ab}}{2},
    \qquad
    |t_J^\triangle| = \frac{E_{ab} - E_b}{6}.
\end{equation}

Thus the triangular approximation estimates the onsite energy and hopping of the {\it graphene-like} bands through a scalar secular equation. The onsite energy is controlled primarily by the two-dimensional confinement inside the triangular core, whereas the bonding--antibonding splitting, and therefore the hopping, is generated by coupling through the connecting trench, which is evanescent only below $\mu_1$.

When both states lie below $\mu_1$ and are strongly localized, $\gamma_0 h\gg1$, the bonding and antibonding boundary coefficients become exponentially close,

\begin{equation}
    \beta_\pm(E) =
    \gamma(E) \left[
    1 \mp 2 e^{-\gamma(E)h} + \mathcal O\!\left(e^{-2\gamma(E)h}\right)
    \right],
\end{equation}
The two levels therefore split around the isolated-junction energy $E_J^{(0)}$, defined by
\begin{equation}
    (3E_J^{(0)}-2\mu_1)
    \sin\!\left(\frac{H\sqrt{E_J^{(0)}}}{2}\right)
    - 3\sqrt{E_J^{(0)}(\mu_1-E_J^{(0)})}
    \cos\!\left(\frac{H\sqrt{E_J^{(0)}}}{2}\right) = 0.
\end{equation}

To leading order, the symmetric shift cancels in the average energy, so that
\begin{equation}
    \epsilon_J^{\triangle} = E_J^{(0)} + \mathcal O\!\left(e^{-2\gamma_0 h}\right),
    \qquad
    \gamma_0=\sqrt{\mu_1-E_J^{(0)}}.
\end{equation}

The hopping is controlled by the exponentially small overlap through the connecting trench. Linearizing the triangular secular equation around $E_J^{(0)}$ gives
\begin{equation}
    |t_J^{\triangle}| =
    A(E_J^{(0)},\mu_1,w)\, e^{-\gamma_0 h} +
    \mathcal O\!\left(e^{-2\gamma_0 h}\right),  
    \qquad
    A =
    \frac{2\gamma_0}{3} \frac{ \dfrac{p_0}{\mu_1} + \dfrac{2p_0}{p_0^2+4\gamma_0^2} }{ \dfrac{H}{4p_0} + \dfrac{1}{2\gamma_0p_0} + \dfrac{\mu_1}{ \gamma_0p_0(p_0^2+4\gamma_0^2)} },
    \qquad
    p_0=\sqrt{E_J^{(0)}}.
\end{equation}

\pagebreak[3]
\section{Dispersive bands of the single channel HQG}
\label{app:singlechannel_dispersive}

For energy-independent vertex phases, the single-channel spectrum repeats
periodically in $q$ in groups of one flat and two generically dispersive
bands. Write $x=(\theta_A-\phi_A)/2$ and $y=(\theta_B-\phi_B)/2$, and use
the honeycomb structure factor
$f(\vk)=1+e^{i\vk\cdot\ve_1}+e^{i\vk\cdot\ve_2}$ of
Appendix~\ref{app:antidot_numerics}. The two dispersive bands satisfy
\begin{equation}\label{dispersive_honeycomb}
 \cos\!\left(2qa_0+\frac{\theta_A+\theta_B+\phi_A+\phi_B}{2}\right)
 =\cos x\cos y+
 \left(1-\frac{2}{9}|f(\vk)|^2\right)\sin x\sin y \, .
\end{equation}
The dispersive bands can be tuned {\it without altering} the flat bands.
Equation~\eqref{dispersive_honeycomb} shows that all bands are flat when
$\sin x\sin y=0$.

For certain values of $(\theta_A,\theta_B,\phi_A,\phi_B)$, the two
dispersive bands form Dirac points at
\begin{equation}
 \vK=\left(\frac{2\pi}{3\sqrt3a_0},\frac{2\pi}{3a_0}\right).
\end{equation}
At $\vK$, the two solutions in each repeated group are
\begin{equation}
 q_{+,n}^{(0)}=\frac{1}{a_0}\left(\pi n-\frac{\theta_B+\phi_A}{2}\right),
 \qquad
 q_{-,n}^{(0)}=\frac{1}{a_0}\left(\pi n-\frac{\theta_A+\phi_B}{2}\right).
\end{equation}
Writing $\vk=\vK+\delta\vk$ and
$q=q_{\sigma,n}^{(0)}+\delta q$, with $\sigma=\pm1$, gives
\begin{equation}\label{eq:dispersive_K_expansion}
 \sigma\sin(x-y)\,a_0\delta q+\cos(x-y)\,(a_0\delta q)^2
 =\frac{a_0^2}{4}|\delta\vk|^2\sin x\sin y
 +O(|a_0\delta\vk|^3,|a_0\delta q|^3).
\end{equation}
For $\sin(x-y)\ne0$ and $\sin x\sin y\ne0$, the leading change in $q$
is quadratic in $|\delta\vk|$. When $x-y=0$ modulo $\pi$ and $\sin y\ne0$, the two
branches have linear dispersion in $q$:
\begin{equation}
 \delta q=\pm \frac{1}{2}|\sin y|\,|\delta\vk|
 +O(a_0|\delta\vk|^2).
\end{equation}
Thus the slope in $q$ is $|\sin y|/2$, reaching $1/2$ only when
$|\sin y|=1$. At a touching with $q_{\sigma,n}^{(0)}\ne0$, the corresponding
energy dispersion is
$E(\vk)=[q_{\sigma,n}^{(0)}]^2
\pm q_{\sigma,n}^{(0)}|\sin y|\,|\delta\vk|
+O(|\delta\vk|^2)$. If $\sin y=0$ as well, the all-flat condition holds
instead.

\section{Multichannel vertex matrices and the parity-mixing obstruction
\label{app:multichannel_parity}}

This appendix derives the most general unitary scattering matrix of a
$D_3$-symmetric multichannel honeycomb vertex in the physical
channel $\otimes$ leg basis.  The essential point is that even- and
odd-parity transverse modes realize the two-dimensional irreducible
representation of $D_3$ in two equivalent but different forms.  The map
between these forms is a nontrivial leg-space intertwiner $J$.  Consequently,
$D_3$ symmetry allows conversion between even and odd transverse channels;
the corresponding blocks of the vertex matrix are proportional to $J$, not
to the identity on the two-dimensional leg sector.  We first obtain the
allowed vertex matrix and then show how the same intertwiner removes the
generic symmetry protection of the single-channel flat band.

\subsection{Physical representation of $D_3$}

We label the three legs at a vertex cyclically by $i=1,2,3$ and use the
channel-major basis $\ket{\mu}\otimes\ket{i}$.  The generators of the
permutation representation on the leg space are chosen as
\begin{equation}
 C=\begin{pmatrix}
 0&1&0\\
 0&0&1\\
 1&0&0
 \end{pmatrix},
 \qquad
 R=\begin{pmatrix}
 0&1&0\\
 1&0&0\\
 0&0&1
 \end{pmatrix}.
 \label{eq:app_D3_generators}
\end{equation}
Here $C$ represents a threefold rotation and $R$ a reflection.  They obey
$C^3=R^2=\mathbb I_3$ and $RCR=C^{-1}$.

Let $N_+$ and $N_-$ denote the numbers of propagating transverse channels
that are respectively even and odd under reflection of an individual wire,
and let $N=N_++N_-$.  We order the channel basis by parity and define
\begin{equation}
 \Pi=\operatorname{diag}(\mathbb I_{N_+},-\mathbb I_{N_-}).
 \label{eq:app_channel_parity}
\end{equation}
A rotation $C_3$ only permutes the three legs, whereas a reflection $M$ also reverses
the transverse coordinate of a wire.  The physical $3N$-dimensional
representation is therefore
\begin{equation}
 \rho(C_3)=\mathbb I_N\otimes C,
 \qquad
 \rho(M)=\Pi\otimes R.
 \label{eq:app_physical_D3_rep}
\end{equation}
Changing the orientation convention of a transverse coordinate can rephase
an odd channel or reverse the sign of the intertwiner introduced below, but
does not change any physical conclusion.

The symmetric leg vector and its orthogonal projectors are
\begin{equation}
 \begin{gathered}
  \ket{e_3}=\frac{1}{\sqrt3}(1,1,1)^T,\\
  P_\parallel=\ket{e_3}\!\bra{e_3},
  \qquad P_\perp=\mathbb I_3-P_\parallel.
 \end{gathered}
 \label{eq:app_leg_projectors}
\end{equation}
The permutation representation decomposes as
$\mathrm{Perm}=\mathbf 1_+\oplus\mathbf 2$, where $\mathbf 1_+$ is
carried by $\operatorname{Im}P_\parallel$ and $\mathbf 2$ by
$\operatorname{Im}P_\perp$.  Including the transverse parity gives
\begin{equation}
 (\mathbf 1_+^{\oplus N_+}\oplus
  \mathbf 1_-^{\oplus N_-})\otimes
 (\mathbf 1_+\oplus\mathbf 2)
 =\mathbf 1_+^{\oplus N_+}\oplus
  \mathbf 1_-^{\oplus N_-}\oplus
  \mathbf 2^{\oplus N}.
 \label{eq:app_D3_decomposition}
\end{equation}
In particular, the even and odd transverse sectors both contain copies of
the same two-dimensional irrep because
$\mathbf 1_-\otimes\mathbf 2\simeq\mathbf 2$.  This equivalence permits the
two parity sectors to mix inside the $P_\perp$ isotypic component.  It does
not identify their physical leg bases.

To make this last distinction explicit, restrict the physical
representation to $\operatorname{Im}P_\perp$.  An even transverse channel
realizes the two-dimensional irrep through
\begin{equation}
 \rho_+(C_3)=C\big|_{\operatorname{Im}P_\perp},
 \qquad
 \rho_+(M)=R\big|_{\operatorname{Im}P_\perp},
 \label{eq:app_even_two_dimensional_rep}
\end{equation}
whereas an odd transverse channel realizes it through
\begin{equation}
 \rho_-(C_3)=C\big|_{\operatorname{Im}P_\perp},
 \qquad
 \rho_-(M)=-R\big|_{\operatorname{Im}P_\perp}.
 \label{eq:app_odd_two_dimensional_rep}
\end{equation}
The two realizations are equivalent, as required by
Eq.~\eqref{eq:app_D3_decomposition}, but their reflection matrices differ by
a sign in the physical basis.  Consequently, a vertex process that converts
an odd channel into an even one must contain a leg-space map $J$ satisfying
\begin{equation}
 J\rho_-(g)=\rho_+(g)J,
 \qquad g\in D_3.
 \label{eq:app_abstract_intertwiner}
\end{equation}
We now construct this map explicitly.

\subsection{The parity intertwiner}

Applying Eq.~\eqref{eq:app_abstract_intertwiner} to the two generators gives
\begin{equation}
 CJ=JC,
 \qquad
 RJ=-JR.
 \label{eq:app_intertwiner_conditions}
\end{equation}
The first relation makes $J$ circulant, so that
$J=\alpha\mathbb I_3+\beta C+\gamma C^2$.  Using
$RCR=C^{-1}=C^2$ in the second relation gives $\alpha=0$ and
$\gamma=-\beta$.  Thus the intertwiner is unique up to an overall scalar.
We choose the normalization
\begin{equation}
 J=\frac{C-C^{-1}}{\sqrt3}
 =\frac{1}{\sqrt3}
 \begin{pmatrix}
 0&1&-1\\
 -1&0&1\\
 1&-1&0
 \end{pmatrix}.
 \label{eq:app_J_definition}
\end{equation}
The normalization is convenient because
\begin{equation}
 \begin{aligned}
  CJ&=JC, & RJ&=-JR,\\
  J^\dagger&=-J, & J^2&=-P_\perp,
  &J^\dagger J&=P_\perp,
 \end{aligned}
 \label{eq:app_J_properties}
\end{equation}
and $JP_\parallel=P_\parallel J=0$.  Thus $J$ is a unitary map on
$\operatorname{Im}P_\perp$ and intertwines the two physical realizations of
$\mathbf 2$.  An equivalent Hermitian convention is
$\mathcal C_\perp=iJ$, for which
$\mathcal C_\perp^2=P_\perp$; this is the convention used in some
representation-theory derivations.

The appearance of $J$ can also be understood as a basis transformation.  In
an abstract isotypic basis, all $N$ copies of $\mathbf 2$ carry the same
two-dimensional matrices, and an equivariant operator has the form
$\mathcal X\otimes\mathbb I_2$.  To return to the physical parity basis, the
odd copies must be transformed by $J$.  On the $P_\perp$ isotypic subspace,
the required map is
\begin{equation}
 \mathcal W=
 \begin{pmatrix}
 \mathbb I_{N_+}\otimes P_\perp&0\\
 0&\mathbb I_{N_-}\otimes J
 \end{pmatrix},
 \qquad
 \mathcal W^\dagger\mathcal W
 =\mathbb I_N\otimes P_\perp.
 \label{eq:app_isotypic_basis_map}
\end{equation}
Removing $J$ from a vertex matrix by using this abstract basis necessarily
transforms the link-propagation matrices as well.  The physical
channel $\otimes$ leg basis keeps both operations explicit.

\subsection{General $D_3$-symmetric unitary vertex matrix}

Let $S_v$ be the scattering matrix at a vertex $v=A,B$.  It maps incoming
propagating amplitudes to outgoing amplitudes and is unitary in a
flux-normalized basis.  We write it in channel-parity blocks, with the even
sector first.  Equations~\eqref{eq:app_physical_D3_rep}--\eqref{eq:app_J_properties}
give the most general matrix satisfying
$[S_v,\rho(C_3)]=[S_v,\rho(M)]=0$:
\begin{equation}
 S_v=
 \begin{pmatrix}
 X_v^{++}\otimes P_\perp+Y_v^+\otimes P_\parallel
 &X_v^{+-}\otimes J\\[1mm]
 X_v^{-+}\otimes J
 &X_v^{--}\otimes P_\perp+Y_v^-\otimes P_\parallel
 \end{pmatrix}.
 \label{eq:app_general_S_physical}
\end{equation}
The superscripts specify the outgoing and incoming channel parities,
respectively.  Thus, for example, $X_v^{+-}$ maps incoming odd channels to
outgoing even channels.  The channel-space dimensions are
\begin{equation}
 \begin{split}
 &X_v^{++}\in\mathbb C^{N_+\times N_+},\qquad
 X_v^{+-}\in\mathbb C^{N_+\times N_-},\\
 &X_v^{-+}\in\mathbb C^{N_-\times N_+},\qquad
 X_v^{--}\in\mathbb C^{N_-\times N_-},
 \end{split}
 \label{eq:app_channel_block_dimensions}
\end{equation}
while $Y_v^+$ and $Y_v^-$ act within the one-dimensional irreps carried
by the symmetric leg sector.  Same-parity scattering is proportional to
$P_\perp$ within the two-dimensional irrep, whereas opposite-parity
scattering carries $J$.  No even--odd block occurs in the $P_\parallel$
sector because $\mathbf 1_+$ and $\mathbf 1_-$ are inequivalent irreps.

The unitarity constraints take a simple form.  Define the multiplicity-space
matrix
\begin{equation}
 \mathsf X_v=
 \begin{pmatrix}
 X_v^{++}&-X_v^{+-}\\
 X_v^{-+}&X_v^{--}
 \end{pmatrix}.
 \label{eq:app_Xmultiplicity}
\end{equation}
Using Eq.~\eqref{eq:app_isotypic_basis_map}, the $P_\perp$ part of
Eq.~\eqref{eq:app_general_S_physical} is
\begin{equation}
 S_{v,\perp}=\mathcal W
 (\mathsf X_v\otimes P_\perp)\mathcal W^\dagger.
 \label{eq:app_Sperp_basis_map}
\end{equation}
It follows that $S_v$ is unitary if and only if
\begin{equation}
 \mathsf X_v\in U(N),
 \qquad
 Y_v^+\in U(N_+),
 \qquad
 Y_v^-\in U(N_-).
 \label{eq:app_unitarity_conditions}
\end{equation}
This form separates symmetry from dynamics: symmetry fixes the leg matrices
$P_\parallel$, $P_\perp$, and $J$, while the unitary matrices in
Eq.~\eqref{eq:app_unitarity_conditions} contain the microscopic scattering
properties of the junction.

If spinless time-reversal symmetry is imposed in a basis of real transverse
modes, reciprocity gives $S_v^T=S_v$.  Since $P_\parallel^T=P_\parallel$,
$P_\perp^T=P_\perp$, and $J^T=-J$, this further implies
\begin{equation}
 \begin{aligned}
  (X_v^{++})^T&=X_v^{++},
  & (X_v^{--})^T&=X_v^{--},\\
  (Y_v^\pm)^T&=Y_v^\pm,
  & X_v^{+-}&=-(X_v^{-+})^T.
 \end{aligned}
 \label{eq:app_reciprocity_constraints}
\end{equation}
\subsubsection{One even and one odd channel}

The smallest parity-mixed example has $N_+=N_-=1$.  In this case
Eq.~\eqref{eq:app_general_S_physical} becomes
\begin{equation}
 S_v=
 \begin{pmatrix}
 x_{v,\perp}^{++}P_\perp+x_{v,\parallel}^{++}P_\parallel
 &x_v^{+-}J\\
 x_v^{-+}J
 &x_{v,\perp}^{--}P_\perp+x_{v,\parallel}^{--}P_\parallel
 \end{pmatrix}.
 \label{eq:app_two_channel_S}
\end{equation}
The symmetric-sector coefficients are phases,
$x_{v,\parallel}^{++}=e^{i\theta_{v,+}}$ and
$x_{v,\parallel}^{--}=e^{i\theta_{v,-}}$.  The remaining four coefficients are
most transparently parametrized through
\begin{equation}
 \begin{pmatrix}
 x_{v,\perp}^{++}&-x_v^{+-}\\
 x_v^{-+}&x_{v,\perp}^{--}
 \end{pmatrix}
 =e^{i\gamma_v}
 \begin{pmatrix}
 e^{i\alpha_v}\cos\eta_v&e^{i\beta_v}\sin\eta_v\\
 -e^{-i\beta_v}\sin\eta_v&e^{-i\alpha_v}\cos\eta_v
 \end{pmatrix}.
 \label{eq:app_U2_parameterization}
\end{equation}
Thus a general $D_3$-symmetric unitary vertex with one channel of each parity
contains six real parameters:
$\theta_{v,+}$, $\theta_{v,-}$, $\gamma_v$, $\alpha_v$, $\beta_v$, and
$\eta_v$.  Explicitly,
\begin{equation}
 \begin{aligned}
 x_{v,\perp}^{++}&=e^{i(\gamma_v+\alpha_v)}\cos\eta_v,
 &x_v^{+-}&=-e^{i(\gamma_v+\beta_v)}\sin\eta_v,\\
 x_v^{-+}&=-e^{i(\gamma_v-\beta_v)}\sin\eta_v,
 &x_{v,\perp}^{--}&=e^{i(\gamma_v-\alpha_v)}\cos\eta_v.
 \end{aligned}
 \label{eq:app_two_channel_coefficients}
\end{equation}
The angle $\eta_v$ controls parity conversion.  The two parity sectors
decouple for $\eta_v=0$.  Spinless reciprocity imposes
$x_v^{+-}=-x_v^{-+}$ and hence, up to channel rephasings,
$\beta_v=\pi/2$ modulo $\pi$.

\subsection{Bloch propagation and the parity-conserving flat band}

We now connect the local result to the honeycomb band problem. For links
of length $a_0$, the channel propagation matrix is
\begin{equation}
 K(E)=\operatorname{diag}
 \bigl(e^{iq_1(E)a_0},\ldots,e^{iq_N(E)a_0}\bigr).
 \label{eq:app_multichannel_propagation}
\end{equation}
For the oriented unit cell used in the main text, the Bloch matrix is
\begin{equation}
 D_{\vk}=\operatorname{diag}
 \left(1,e^{-i\vk\cdot\mathbf e_1},
 e^{-i\vk\cdot\mathbf e_2}\right).
 \label{eq:app_Dk}
\end{equation}
One round trip from an $A$ vertex to a $B$ vertex and back is described by
\begin{equation}
 U(E,\vk)=
 (K\otimes D_{\vk}^\dagger)S_B
 (K\otimes D_{\vk})S_A,
 \label{eq:app_multichannel_roundtrip}
\end{equation}
and the Bloch energies obey
\begin{equation}
 \det[\mathbb I_{3N}-U(E,\vk)]=0.
 \label{eq:app_multichannel_secular}
\end{equation}

The single-channel destructive-interference construction is encoded by a
one-dimensional leg subspace at generic momentum.  Write
\begin{equation}
 a_{\vk}=e^{-i\vk\cdot\mathbf e_1},
 \qquad
 b_{\vk}=e^{-i\vk\cdot\mathbf e_2},
 \label{eq:app_abk}
\end{equation}
and define
\begin{equation}
 \ket{r_{\vk}}=
 \begin{pmatrix}
 b_{\vk}-a_{\vk}\\
 1-b_{\vk}\\
 a_{\vk}-1
 \end{pmatrix}.
 \label{eq:app_flat_leg_vector}
\end{equation}
For $\vk\ne\Gamma$, this is the unnormalized version of the
single-channel vector in Eq.~\eqref{eq:alphak_singlechannel}:
$\ket{r_{\vk}}=\sqrt{N_{\vk}}\ket{\alpha_{\vk}}$. It satisfies
\begin{equation}
 P_\perp\ket{r_{\vk}}=\ket{r_{\vk}},
 \qquad
 P_\perp D_{\vk}\ket{r_{\vk}}
 =D_{\vk}\ket{r_{\vk}}.
 \label{eq:app_flat_leg_conditions}
\end{equation}
At $\Gamma$, $\ket{r_{\vk}}$ vanishes and the limiting null space is
the full two-dimensional space $\operatorname{Im}P_\perp$.

Suppose first that transverse parity is conserved at both vertices,
\begin{equation}
 X_A^{+-}=X_A^{-+}=X_B^{+-}=X_B^{-+}=0.
 \label{eq:app_no_parity_mixing}
\end{equation}
The two parity sectors then separate.  In either sector $p=+$ or $-$, let
$K_p(E)$ be the restriction of $K(E)$ to that channel space.  For a channel
vector $\ket{\phi_p}$,
Eqs.~\eqref{eq:app_multichannel_roundtrip}
and \eqref{eq:app_flat_leg_conditions} give
\begin{equation}
 U(E,\vk)
 [\ket{\phi_p}\otimes\ket{r_{\vk}}]
 =
 [K_pX_B^{pp}K_pX_A^{pp}\ket{\phi_p}]
 \otimes\ket{r_{\vk}}.
 \label{eq:app_parity_sector_reduction}
\end{equation}
Every solution of
\begin{equation}
 \det[K_p(E)X_B^{pp}K_p(E)X_A^{pp}-\mathbb I_{N_p}]=0
 \label{eq:app_parity_sector_FB_condition}
\end{equation}
therefore produces an exact flat band: the energy condition contains no
Bloch momentum.  The multichannel flat-band construction remains valid
within a single parity sector, including when several channels of that same
parity mix.

\subsection{Even--odd mixing and flat-band protection}

When the opposite-parity blocks are present, local scattering sends the leg
vector $\ket{r_{\vk}}$ to $J\ket{r_{\vk}}$ during a parity
conversion.  Although $J\ket{r_{\vk}}$ remains in
$\operatorname{Im}P_\perp$ at the vertex, propagation to the neighboring
vertex does not preserve the destructive-interference condition.  Indeed,
direct substitution of Eqs.~\eqref{eq:app_J_definition},
\eqref{eq:app_Dk}, and \eqref{eq:app_flat_leg_vector} gives
\begin{equation}
 \bra{e_3}D_{\vk}J\ket{r_{\vk}}
 =\frac{2}{3}\left(
 1-a_{\vk}-b_{\vk}
 +a_{\vk}^2+b_{\vk}^2
 -a_{\vk}b_{\vk}\right).
 \label{eq:app_J_leakage}
\end{equation}
The right-hand side is nonzero at generic $\vk$.  The converted
amplitude therefore reaches the next vertex with a component in
$\operatorname{Im}P_\parallel$.  Because the vertex matrix acts differently on the $P_\parallel$ and
$P_\perp$ components, the reduction in
Eq.~\eqref{eq:app_parity_sector_reduction} no longer applies in general.

The same obstruction appears in the isotypic basis of
Eq.~\eqref{eq:app_isotypic_basis_map}, where the local $P_\perp$ scattering
has the form $\mathsf X_v\otimes\mathbb I_2$. The
$P_\perp$-to-$P_\perp$ propagation blocks are
$P_\perp D_{\vk}P_\perp$ for even channels and
$J^\dagger D_{\vk}J$ for odd channels. They differ at generic
$\vk$: the basis change that gives the vertex scattering a common
leg-space form makes propagation depend on transverse parity.

The construction above therefore no longer guarantees a flat band
when even and odd channels mix, even in a $D_3$-symmetric system. Exact
flatness can nevertheless survive under additional restrictions, for example
when
\begin{enumerate}
 \item only one transverse-parity sector is present in the energy window;
 \item an additional symmetry enforces Eq.~\eqref{eq:app_no_parity_mixing};
 \item the relevant channel vector lies in the common kernel of all
 opposite-parity blocks encountered in a round trip; or
 \item microscopic parameters are fine-tuned so that the momentum-dependent
 contributions cancel.
\end{enumerate}
For generic parity-mixed scattering, the $J$ blocks couple the
destructive-interference sector to the symmetric leg sector and give the band
a finite width.

\subsection{Weak parity mixing}

When parity conversion is weak, its leading effect can be stated without
choosing a microscopic junction model.  Write the round-trip matrix of Eq.~\eqref{eq:app_multichannel_roundtrip}
as
\begin{equation}
 U(E,\vk;\varepsilon)
 =U^{(0)}(E,\vk)
 +\varepsilon U^{(1)}(E,\vk)
 +\varepsilon^2U^{(2)}(E,\vk)+\cdots,
 \label{eq:app_U_mixing_expansion}
\end{equation}
where the opposite-parity blocks $X_v^{+-}$ and $X_v^{-+}$ are of order
$\varepsilon$. The matrix coefficients follow by expanding the
chosen vertex matrices in Eq.~\eqref{eq:app_multichannel_roundtrip}; in
particular, $U^{(2)}=\frac{1}{2}\left.\partial_\varepsilon^2U\right|_{\varepsilon=0}$
at fixed $E$ and $\vk$. We vary only the parity-conversion strength, so the
parity-preserving blocks have no linear correction and $U^{(1)}$ changes
parity. Let a flat band in parity sector $p=\pm1$ be nondegenerate away from
its symmetry-enforced touching at $\Gamma$, and let it satisfy
$U^{(0)}(E_\star,\vk)\ket{R_{\vk}}=\ket{R_{\vk}}$, with
left eigenvector $\bra{L_{\vk}}$ normalized by
$\braket{L_{\vk}|R_{\vk}}=1$.  A single insertion of
$U^{(1)}$ changes transverse parity, so
\begin{equation}
 \bra{L_{\vk}}U^{(1)}\ket{R_{\vk}}=0.
 \label{eq:app_first_order_zero}
\end{equation}
At the fixed unperturbed energy $E_\star$, we therefore define the
eigenvalue correction coefficient by

\begin{equation*}
 \lambda(E_\star,\vk;\varepsilon)
 =1+\varepsilon^2\delta\lambda^{(2)}(\vk)+O(\varepsilon^3).
\end{equation*}

All matrix coefficients below are evaluated at $(E_\star,\vk)$.
The eigenvector also changes:
$\ket{R_{\vk}(\varepsilon)}=\ket{R_{\vk}}
+\varepsilon\ket{R_{\vk}^{(1)}}+O(\varepsilon^2)$.
Choosing $\braket{L_{\vk}|R_{\vk}^{(1)}}=0$, the second-order
eigenvalue equation projected onto $\bra{L_{\vk}}$ gives
\begin{equation*}
 \delta\lambda^{(2)}(\vk)
 =\bra{L_{\vk}}U^{(2)}\ket{R_{\vk}}
 +\bra{L_{\vk}}U^{(1)}\ket{R_{\vk}^{(1)}}.
\end{equation*}
The first term is the direct contribution of $U^{(2)}$; the second
comes from the eigenvector change driven by $U^{(1)}$.
The first-order equation
$(\mathbb I-U^{(0)})\ket{R_{\vk}^{(1)}}=U^{(1)}\ket{R_{\vk}}$
places this correction in the opposite-parity sector, whose projector is
\begin{equation*}
 \mathcal Q_{-p}=\frac{\mathbb I_N-p\Pi}{2}\otimes\mathbb I_3.
\end{equation*}
Here $\Pi$ is the channel-parity matrix defined in
Eq.~\eqref{eq:app_channel_parity}. The projector retains all channels of
parity $-p$ on all three legs and is independent of $\vk$ in this
basis.
Away from a degeneracy with the opposite-parity spectrum, solving
this first-order equation in that sector and substituting above gives
\begin{equation}
 \begin{split}
 \delta\lambda^{(2)}(\vk)={}&
 \bra{L_{\vk}}U^{(2)}\ket{R_{\vk}}\\
 &+\bra{L_{\vk}}U^{(1)}\mathcal Q_{-p}
 [\mathbb I-\mathcal Q_{-p}U^{(0)}\mathcal Q_{-p}]^{-1}\\
 &\hspace{8mm}\times
 \mathcal Q_{-p}U^{(1)}\ket{R_{\vk}},
 \end{split}
 \label{eq:app_second_order_mixing}
\end{equation}
The resolvent term describes two parity conversions,
$p\to-p\to p$. For an isolated eigenvalue, this restriction gives the same
correction as the usual complementary projector
$\mathbb I_{3N}-\ket{R_{\vk}}\!\bra{L_{\vk}}$, because
$U^{(1)}\ket{R_{\vk}}$ has parity $-p$. At $\Gamma$, removing only
one state would leave its degenerate same-parity partner in the resolvent
and make the inverse singular; $\mathcal Q_{-p}$ excludes that partner.
The touching doublet must still be treated as a degenerate subspace.
Within it, both the second-order correction and the projected energy
derivative are scalar by $D_3$ symmetry, so either normalized vector gives
the same energy shift. We assume no additional degeneracy and
$\partial_E\lambda_0(E_\star)\ne0$.
The shifted band energy must restore the eigenvalue to one, giving
\begin{equation}
 \delta E(\vk)
 =-\varepsilon^2
 \frac{\delta\lambda^{(2)}(\vk)}{\partial_E\lambda_0(E_\star)},
 \label{eq:app_energy_shift_mixing}
\end{equation}
where $\lambda_0(E)$ is the unperturbed round-trip eigenvalue that crosses
one at $E_\star$.  The matrix elements in
Eq.~\eqref{eq:app_second_order_mixing} contain the nonconstant geometric
factor generated by $D_{\vk}$ and $J$, exemplified by
Eq.~\eqref{eq:app_J_leakage}; hence the correction is generically dispersive.
If an even and an odd unperturbed band are degenerate, the resolvent in
Eq.~\eqref{eq:app_second_order_mixing} is singular and degenerate
perturbation theory can give a splitting already at order $\varepsilon$.

For Fig.~\ref{fig:multichannel_parity_mixing}, we solve for zero
eigenphases of $U(E,\vk;\eta)$ and continue each descendant from
$\eta=0$ using eigenvector overlaps (subspace overlaps at degeneracies).
The bandwidth extrema are evaluated on $30\times30$ and $60\times60$
uniform grids in the Bloch phases $\vk\cdot\ve_{1,2}$, including
$\Gamma$ and $K$; the resulting widths agree within $10^{-8}E_0$.

The local symmetry analysis therefore identifies both the protected limit
and its generic obstruction.  Same-parity multichannel scattering preserves
the flat-band reduction in Eq.~\eqref{eq:app_parity_sector_FB_condition}.
Opposite-parity scattering is symmetry allowed, carries the intertwiner $J$,
and produces momentum-dependent coupling after propagation through the honeycomb links.

\clearpage
\twocolumngrid

\bibliography{HdrThesis.bib}

\end{document}